\documentclass[article]{aa}
\usepackage{graphicx}
\usepackage[varg]{txfonts}
\usepackage[dvipsnames]{xcolor}
\usepackage{hyperref}
\hypersetup{pdfborder = {0 0 0},
    colorlinks = true,
    linkbordercolor = {white},
    citecolor=NavyBlue,
    urlcolor=Green
}

\newcommand{\orcidlink}[1]{\protect\href{https://orcid.org/#1}{\protect\includegraphics[width=8pt]{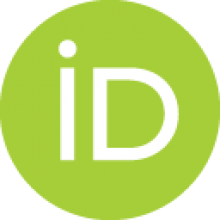}}}

\usepackage{amssymb}
\usepackage{amsmath}
\usepackage{xspace}
\usepackage{array,multirow}
\usepackage{enumitem}
\usepackage[english]{babel}
\usepackage{mathrsfs}
\usepackage{listings}
\usepackage{blindtext}
\usepackage{lipsum} 
\usepackage{subfigure}
\usepackage[T1]{fontenc}
\usepackage[utf8]{inputenc}
\usepackage{ae,aecompl}
\usepackage{caption}
\usepackage{threeparttable}
\usepackage{booktabs}

\usepackage{natbib,twoopt}

\newcommand{\ie}{i.e.\@\xspace} 
\newcommand{\eg}{e.g.\@\xspace} 

\renewcommand{\arraystretch}{1.}

\renewcommand{\eqref}[1]{Eq.~\ref{#1}}
\newcommand{\fref}[1]{Fig.~\ref{#1}}
\newcommand{\Fref}[1]{Figure~\ref{#1}}
\newcommand{\tref}[1]{Table~\ref{#1}}
\newcommand{\sref}[1]{Sect.~\ref{#1}}
\newcommand{\Sref}[1]{Section~\ref{#1}}
\newcommand{\aref}[1]{Appendix~\ref{#1}}

\newcommand{\numax}{\ensuremath{\nu_{\rm max}}\xspace}
\newcommand{\dnu}{\ensuremath{\Delta\nu}\xspace}

\newcommand{\muhz}{\ensuremath{\mu\rm Hz}\xspace}

\newcommand{\kp}{\emph{Kepler}\xspace}

\newcommand{\Kp}{\ensuremath{\rm Kp}\xspace}
\newcommand{\teff}{\ensuremath{T_{\rm eff}}\xspace}

\newcommand{\logg}{\ensuremath{\log g}\xspace}
\newcommand{\feh}{\ensuremath{\rm [Fe/H]}\xspace}
\newcommand{\vsini}{\ensuremath{v\sin i}\xspace}

\numberwithin{equation}{section}

\makeatletter
\def\maketag@@@#1{\hbox{\m@th\normalfont\normalsize#1}}
\makeatother

\bibpunct{(}{)}{;}{a}{}{,} 

\makeatletter
\newcommand*\mysize{%
  \@setfontsize\mysize{5.7}{8.0}%
}
\makeatother

\makeatletter
\newcommand*\tabsize{%
  \@setfontsize\tabsize{7.}{8.0}%
}
\makeatother

\makeatletter
\newcommand\footnoteref[1]{\protected@xdef\@thefnmark{\ref{#1}}\@footnotemark}
\makeatother

\begin{document}

\title{The K2 Asteroseismic KEYSTONE sample of Dwarf and Subgiant Solar-Like Oscillators}
\subtitle{I: Data and Asteroseismic parameters}

\titlerunning{The K2 asteroseismic sample. I.}

\author{
Mikkel N. Lund\inst{\ref{I1}}\orcidlink{0000-0001-9214-5642} \and 
Sarbani Basu\inst{\ref{I2}}\orcidlink{0000-0002-6163-3472} \and 
Allyson~Bieryla\inst{\ref{I3}}\orcidlink{0000-0001-6637-5401} \and 
Luca Casagrande\inst{\ref{I4}}\orcidlink{0000-0003-2688-7511} \and 
Daniel Huber\inst{\ref{I5},\ref{I6}}\orcidlink{0000-0001-8832-4488}\and 
Saskia Hekker\inst{\ref{I7},\ref{I8},\ref{I9},\ref{I1}}\orcidlink{0000-0002-1463-726X} \and 
Lucas Viani\inst{\ref{I2}} \and 
Guy~R.~Davies\inst{\ref{I10},\ref{I1}}\orcidlink{0000-0002-4290-7351} \and 
Tiago~L.~Campante\inst{\ref{I11},\ref{I12}}\orcidlink{0000-0002-4588-5389} \and 
William~J.~Chaplin\inst{\ref{I10},\ref{I1}}\orcidlink{0000-0002-5714-8618} \and 
Aldo M. Serenelli\inst{\ref{I13},\ref{I14}}\orcidlink{0000-0001-6359-2769} \and 
J. M. Joel Ong\inst{\ref{I2},\ref{I5}}\thanks{Hubble Fellow}\orcidlink{0000-0001-7664-648X} \and 
Warrick H. Ball\inst{\ref{I15},\ref{I10},\ref{I1}}\orcidlink{0000-0002-4773-1017} \and 
Amalie Stokholm\inst{\ref{I10},\ref{I16},\ref{I17}}\orcidlink{0000-0002-5496-365X} \and 
Earl P. Bellinger\inst{\ref{I2}}\orcidlink{0000-0003-4456-4863} \and 
Micha\"{e}l~Bazot\inst{\ref{I8}}\orcidlink{0000-0003-0166-1540} \and
Dennis~Stello\inst{\ref{dennis1},\ref{I6},\ref{dennis3}}\orcidlink{0000-0002-4879-3519} \and  
David~W.~Latham\inst{\ref{I3}}\orcidlink{0000-0001-9911-7388} \and 
Timothy~R.~White\inst{\ref{I6},\ref{I1}}\orcidlink{0000-0002-6980-3392} \and 
Maryum~Sayeed\inst{\ref{I18}}\orcidlink{0000-0001-6180-8482} \and 
V{\'{\i}}ctor~Aguirre~B{\o}rsen-Koch\inst{\ref{I19}}\orcidlink{0000-0002-6137-903X} \and 
Ashley Chontos\inst{\ref{I5}}\orcidlink{0000-0003-1125-2564}
   }
\offprints{MNL, \email{mikkelnl@phys.au.dk}}          

\institute{
Stellar Astrophysics Centre, Department of Physics and Astronomy, Aarhus University, Ny Munkegade 120, DK-8000 Aarhus C, Denmark\label{I1} \and
Department of Astronomy, Yale University, PO Box 208101, New Haven, CT 06520-8101, USA\label{I2} \and
Center for Astrophysics | Harvard-Smithsonian, 60 Garden Street Cambridge, MA 02138 USA\label{I3} \and
Research School of Astronomy and Astrophysics, Mount Stromlo Observatory, The Australian National University, ACT 2611, Australia\label{I4} \and
Institute for Astronomy, University of Hawai`i, 2680 Woodlawn Drive, Honolulu, HI 96822, USA\label{I5} \and
Sydney Institute for Astronomy (SIfA), School of Physics, University of Sydney, NSW 2006, Australia\label{I6} \and
Center for Astronomy (ZAH/LSW), Heidelberg University, K\"{o}nigstuhl 12, 69117 Heidelberg, Germany\label{I7} \and
Heidelberger Institut f\"{u}r Theoretische Studien, Schloss-Wolfsbrunnenweg 35, 69118 Heidelberg, Germany\label{I8} \and
Max Planck Institute for Solar System Research, G{\"o}ttingen, Germany\label{I9} \and
School of Physics and Astronomy, University of Birmingham, Edgbaston, Birmingham, B15 2TT, UK\label{I10} \and
Instituto de Astrof\'{i}sica e Ci\^{e}ncias do Espa\c{c}o, Universidade do Porto, Rua das Estrelas, 4150-762 Porto, Portugal\label{I11} \and
Departamento de F\'{i}sica e Astronomia, Faculdade de Ci\^{e}ncias da Universidade do Porto, Rua do Campo Alegre, s/n, 4169-007 Porto, Portugal\label{I12} \and
Institute of Space Sciences (ICE, CSIC) Campus UAB, Carrer de Can Magrans, s/n, 08193, Bellaterra, Spain\label{I13} \and
Institut d'Estudis Espacials de Catalunya (IEEC), C/Gran Capita, 2-4, 08034, Barcelona, Spain\label{I14} \and
Advanced Research Computing, University of Birmingham, Edgbaston, Birmingham, B15 2TT, UK\label{I15} \and
Dipartimento di Fisica e Astronomia, Universit\`{a} degli Studi di Bologna, Via Gobetti 93/2, I-40129 Bologna, Italy\label{I16} \and
INAF - Osservatorio di Astrofisica e Scienza dello Spazio di Bologna, Via Gobetti 93/3, I-40129 Bologna, Italy\label{I17} \and
School of Physics, University of New South Wales, NSW 2052, Australia\label{dennis1} \and
The Australian Research Council Centre of Excellence for All Sky Astrophysics in 3 Dimensions (ASTRO 3D), Australia\label{dennis3} \and
Department of Astronomy, Columbia University, 550 West 120th Street, New York, NY, USA\label{I18} \and
DARK, Niels Bohr Institute, University of Copenhagen, Jagtvej 128, 2200, Copenhagen, Denmark\label{I19}
}              

\authorrunning{Lund et al.}

\date{Received: 21 March 2024; Accepted: 20 May 2024}

\abstract
  {}
  {The KEYSTONE project aims to enhance our understanding of solar-like oscillators by delivering a catalogue of global asteroseismic parameters (\dnu\ and \numax) for 173 stars, comprising mainly dwarfs and subgiants, observed by the K2 mission in its short-cadence mode during campaigns 6-19.} 
  {We derive atmospheric parameters and luminosities using spectroscopic data from TRES, astrometric data from \textit{Gaia}, and the infrared flux method (IRFM) for a comprehensive stellar characterisation. Asteroseismic parameters are robustly extracted using three independent methods, complemented by an iterative refinement of the spectroscopic analyses using seismic \logg\ values to enhance parameter accuracy.}
  {Our analysis identifies new detections of solar-like oscillations in 159 stars, providing an important complement to already published results from previous campaigns. The catalogue provides homogeneously derived atmospheric parameters and luminosities for the majority of the sample. Comparison between spectroscopic \teff and those obtained from the IRFM demonstrates excellent agreement. The iterative approach to spectroscopic analysis significantly enhances the accuracy of the stellar properties derived.}
  {}


\maketitle


\section{Introduction} \label{sec:intro}

For the last decade and a half, the advent of space-based photometric missions has ushered in a new era of precision stellar astrophysics from the utilisation of asteroseismology. Starting with CoRoT \citep{Auvergne2009,Michel2008,DeRidder2009} and \kp \citep{Gilliland2010}, followed by K$2$ \citep{K2Howell}, and currently with the ongoing observations of TESS \citep{Ricker2014}, these missions provide the required observational ingredients for studying the internal resonant oscillations of stars \citep{Aerts2010,garcia2019LRSP}. By probing the stellar interior, asteroseismology has a unique capability of providing precise stellar parameters, in particular the mean density ($\langle\rho\rangle$), surface gravity (\logg), mass ($M$), radius ($R$), and age ($\tau$).  

To date, the \kp/K2 missions have delivered the main basis for such analysis, with stellar parameter catalogues based on global seismic parameters and spectroscopic information.  For red giants the most notable are the APOKASC \citep{Pinsonneault2014,Pinsonneault2018} and APO-K2 samples \citep{Zinn2022,Stasik2023}.
For main-sequence (MS) and sub-giant (SG) stars \citet{Chap2014} provided the first comprehensive catalogue of stellar parameters from global seismic parameters, which was augmented with homogeneous spectroscopic inputs by \citet{serenelli2017}, and by additional detections by \citet{Balona2020} and \citet{Mathur2021}. 

\begin{figure}
    \centering
    \includegraphics[width=1\columnwidth]{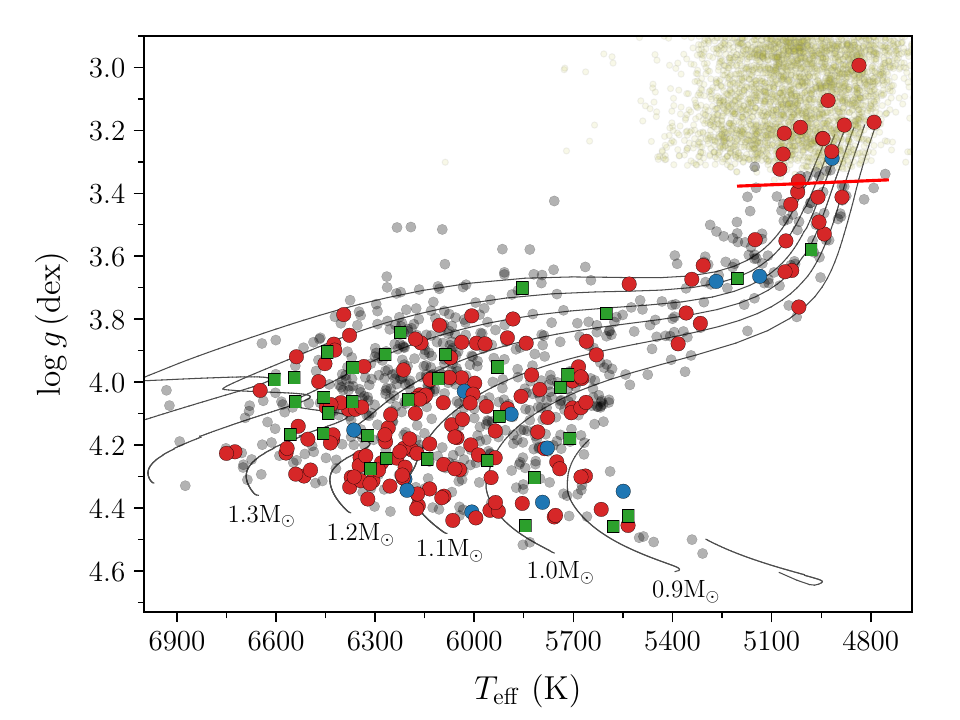}
    \caption{Kiel-diagram of the $173$ stars from C6-19 with detected oscillations analysed in this study, in addition to the $33$ C1-3 stars of \citet{k2chap} and \citet{keystone}. Red circular markers indicate stars with \logg and \teff values from our spectroscopic analysis (\sref{sec:spec}), while blue circles indicate stars where these were only available from the IRFM (\sref{sec:irfm}). The green squares indicate the SPC values for the C1-3 stars \citep{keystone}. The $625$ solar-like oscillators from \kp SC data are shown with smaller gray background markers, using \logg and \teff from \citet{Mathur2021}. The yellow markers to the upper right show the high-\logg part of the \citet{Yu2018} sample of \kp giants. The full red line marks a \numax equal to the \kp LC Nyquist frequency of ${\sim}283\rm \, \mu Hz$. Evolutionary tracks were calculated using GARSTEC \citep{Weiss2008} with $\feh=0$.}
    \label{fig:kiel}
\end{figure}

In this paper, we introduce the first part of the KEYSTONE catalogue of stellar parameters for solar-like MS/SG oscillators observed by the K2 mission in its short-cadence (SC; $\delta t{\sim}1$ min) mode. Our analysis focuses on measuring the sample's global asteroseismic and stellar atmospheric parameters. A second paper (hereafter referred to as Paper~II) will provide results on the stellar modelling. This work builds on earlier catalogues from the initial K2 campaigns (C) 1-3 by \citet{k2chap} and \citet{keystone}, as well as cluster studies from C4-5 data by \citet{Stello2016} and \citet{hyades}, extending them to encompass the entire K2 mission up to C19. Results are presented for $173$ stars with detected \numax, the frequency of maximum oscillation power, and \dnu, the mean large frequency separation, from C6-19. This includes $159$ new detections and a homogeneous set of spectroscopic observations for $163$ of the stars. The targets of the KEYSTONE project are shown in a Kiel-diagram\footnote{The term ``\textit{Kiel-diagram}'' appears to have been used first by \citet{1983QJRAS..24..393C} about diagrams introduced by members of the astronomy group at Kiel University \citep[see, \eg,][fig. 12]{1955ZA.....36...42H} (Charles Cowley, private communication)} in \fref{fig:kiel}, alongside known targets from \kp \citep{Mathur2021,Yu2018}. When combined with earlier detections and analyses from C1-5, the total KEYSTONE sample of $210$ stars significantly augments the existing collection of $625$ solar-like MS/SG oscillators from the \kp mission \citep{Chap2014, serenelli2017, Balona2020, Mathur2021}.

The paper is structured as follows: \sref{sec:select} describes the target selection, while \sref{sec:data} outlines the input data for our analysis. \Sref{sec:stellarparams} describes our analysis of the provided stellar parameters, including atmospheric parameters in \sref{sec:atmos}, luminosities in \sref{sec:lum}, and asteroseismic parameters in \sref{sec:ana}. We conclude and provide an outlook in \sref{sec:con}. 


\renewcommand{\arraystretch}{1.05} 

\begin{table*}
\caption{Number of targets associated with the KEYSTONE study.}
\label{tab:fut_cam}
\centering 
\begin{tabular}{lcccccc}
\toprule
Cam. & $\#$ targets & $\#$ detections & Success rate (\%) & Proposal\tablefootmark{a} & PI\tablefootmark{b} & Notes \\
\midrule \addlinespace[3pt]
1 & 24 & 4 & 17 & 1038 & Chaplin   & High noise\\
2 & 33 & 5  & 15 & 2023 & Chaplin   & High noise\\
3 & 33 & 24 & 73 & 3023 & Chaplin  & South Galactic Cap\\
4 & 31 & 2\tablefootmark{c} & 6 & 4074 & Basu  & Hyades/Pleiades\\
5 & 51 & 6\tablefootmark{d} & 12 & 5074 & Basu  & M44/M67\\
6 & 35 & 22 & 63 & 6039 & Davies  & North Galactic cap\\
7 & 17 & 8 & 47 & 7039 & Davies  & Near galactic centre\\
8 & 10 & 5 & 50 & 8002 & Campante  &\\
9 & 0 & 0 & ---  & ---  &  --- &  Galactic centre\\
10 & 35 & 13 & 37 & 10002 & Campante &  North Galactic cap\\
11 & 28 & 18 & 64 & 11012 & Lund &  Galactic centre\\
12 & 37 & 24 & 65 & 12012 & Lund &  South Galactic cap\\
13 & 38 & 12 & 32 & 13012 & Lund &  Hyades\\
14 & 46 & 27 & 59 & 14010 & Lund &  North Galactic cap\\
15 & 45 & 24 & 53 & 15010 & Lund &  \\
16 & 31 & 7 & 23 & 16010 & Lund & M44/M67 \\
17 & 15 & 11 & 73 & 17036 & Lund &  \\
18 & 21 & 10 & 48 & 18036 & Lund & M44/M67 \\
19 & 16 & 10 & 62 & 19036 & Lund &  \\
\midrule
 & 546 (492) & 232 (210\tablefootmark{e}) & 42 (43) & & & \\
\bottomrule
\end{tabular}
\tablefoot{Overview of the number of targets associated with the KEYSTONE study that were observed in SC (not counting if only LC observations were obtained) in the different campaigns, together with the number of detections made. The bottom row provides sums of the columns, with values for the number of unique targets in parenthesis.
\tablefoottext{a}{Proposal ID within the K2 guest observer (GO) program}
\tablefoottext{b}{Principal investigator}
\tablefoottext{c}{Hyades analysis by \citet{hyades}}
\tablefoottext{d}{M67 analysis by \citet{Stello2016}}
\tablefoottext{e}{three of these detections were only possible from combining several campaigns.}
}
\end{table*}
\renewcommand{\arraystretch}{1} 

\section{The sample}\label{sec:select}
Stars observed for this study cover C1-19 and were proposed via the K2 guest observer program (see \tref{tab:fut_cam}). Results from C1-3 have been presented in \citet{k2chap} and \citet{keystone}. We note that C4-5 were dedicated to identifying solar-like oscillators in the open clusters M44, Hyades, and M67. Results from these observations have been presented in \citet{hyades} (Hyades) and \citealt{Stello2016} (M67), and will not be re-analysed in this study. Hence in this work, we focus on the analysis of stars from C6-19. Some M67 C5 stars were re-observed in C16 and 18, and we will provide independent results from these latter campaigns. No targets were proposed in C9 as this campaign targeted the galactic bulge mainly for microlensing observations \citep[see, \eg,][]{Kim2018}.

\begin{figure*}
    \centering
    \includegraphics[width=\textwidth]{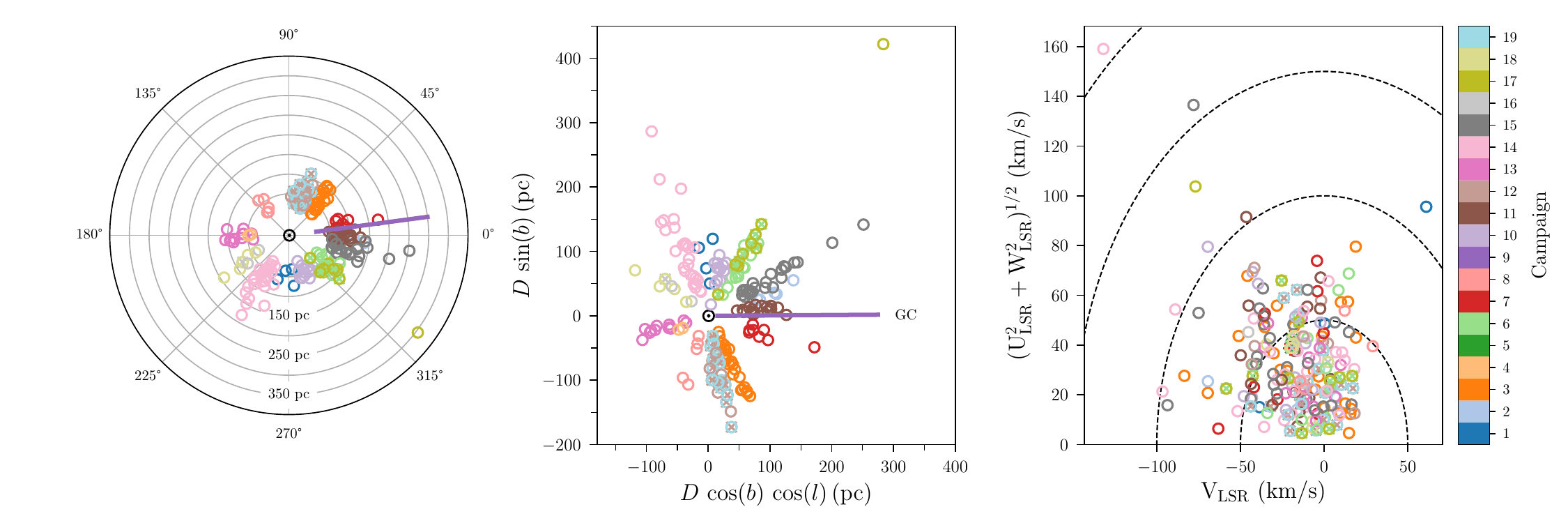}
    \caption{Sky positions and velocities in galactic coordinates of targets observed in C1-19 with positive seismic detections. We have generally adopted the photogeometric distances from \citet{BJ2021}, radial velocities from our SPC analysis (\sref{sec:spec}), and proper motions from \textit{Gaia} EDR3 \citep{EDR3_2021}. For targets with observations in more than one campaign, the targets will be indicated by crosses for one of these on top of the corresponding circular marker from the other observation campaign. We note that M67 targets, at a distance of ${\sim}800$ pc from the Sun have been omitted from the figure. Left: Positions of targets in galactic longitude ($l$) and distance ($D$) from the Sun. The different colours indicate the K2 campaign (see colour bar in right panel). For C9, where no targets were proposed, we have indicated the direction with the coloured line. The galactic centre (GC) is in the direction of $l = 0^{\circ}$. Middle: Positions projected in the abscissa onto the $l = 180^{\circ} \rightarrow 0^{\circ}$ line, with the direction of the GC to the right. Here $b$ denotes the galactic latitude. Right: distribution of galactic velocities, using a local standard of rest (LSR) of $\rm (U, V, W)=(8.63, 4.76, 7.26)\,\, km/s$ \citep{Ding_2019}. Dashed circles indicate total velocities in steps of $\rm 50\, km/s$.}
    \label{fig:sky}
\end{figure*}

The targets proposed for observations in a given campaign were selected based on a predicted detectability of solar-like oscillations \citep[see][]{2011ApJ...732...54C,keystone} and a \numax above the Nyquist frequency of ${\sim}283\,\rm\mu Hz$ for long-cadence (LC) observations. In addition to the detectability the target selection included a prioritization based on the stellar brightness, the relative uncertainty on parallax, and the proximity to detector edges and other bright targets; targets nearer the center of the field were given higher priority since they cost less in pixels and are generally slightly less noisy due to the reduced effects of the spacecraft roll \citep{2016van.cleve.pasp, keystone}. To promote interesting science cases we finally adjusted the rankings based on existing information on the stars, \eg, cluster membership, known exoplanets, etc.
In some campaigns covering known exoplanet hosts, these were included despite a low predicted detectability of solar-like oscillations -- we refer to Chontos et al. (in prep.) for an in-depth analysis of the exoplanet systems with seismic hosts.
\begin{figure*}
    \centering
    \includegraphics[width=\textwidth]{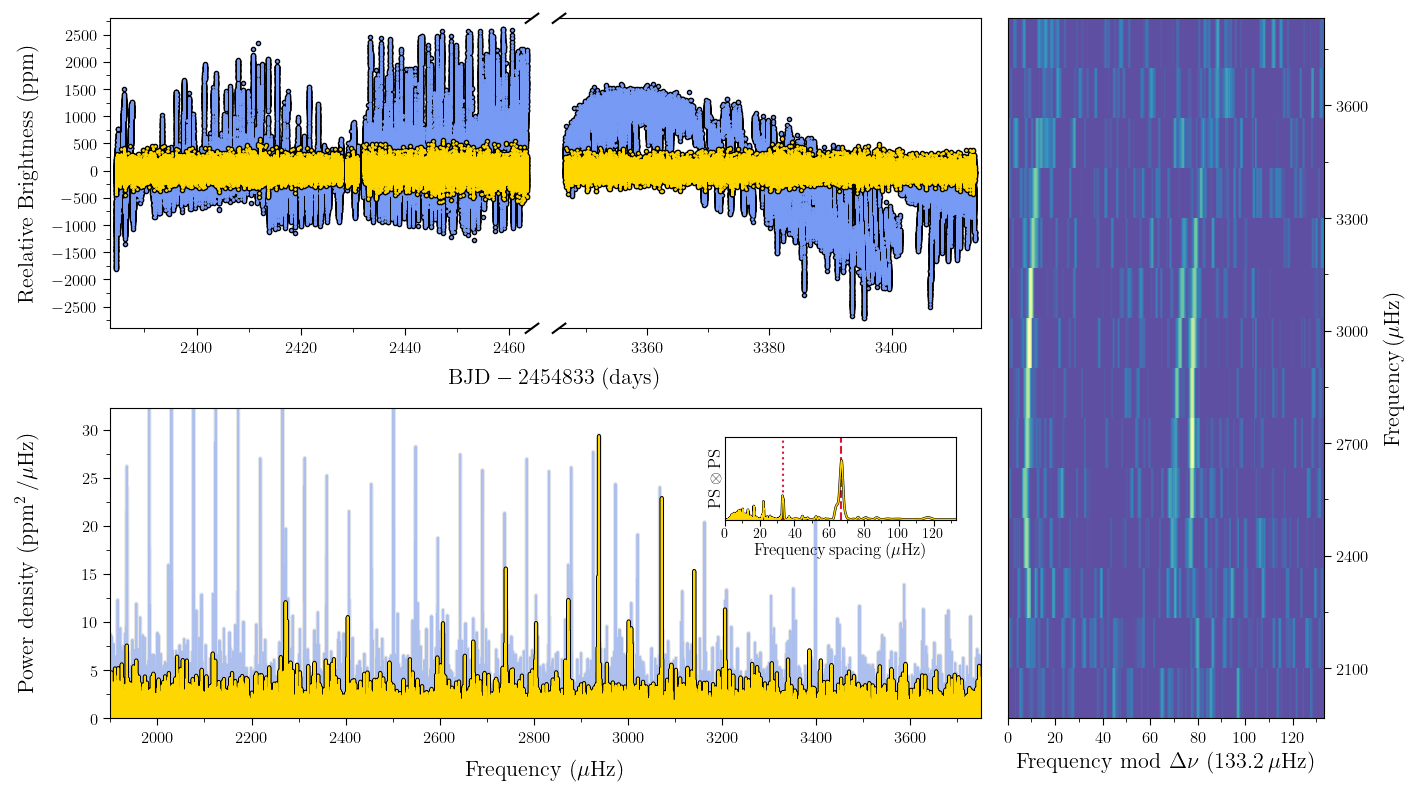}
    \caption{K2 photometry processing example for \object{EPIC 212708252}. Top left: light curve for \object{EPIC 212708252} obtained during K2 C6 (left part) and C17 (right part). Blue points show the raw light curve as extracted from the target pixel files using custom apertures, while yellow points show the light curve after correcting for the K2 systematics. Bottom left: Power density spectra for \object{EPIC 212708252} as calculated from the raw (blue) and systematics-corrected (yellow) light curves. The insert shows the $\rm PS\otimes PS$ of a region of the PDS centred on the measured \numax ($\rm{\sim}2900\, \mu Hz$), where the dashed (dotted) line corresponds to the measured value for $\dnu/2$ ($\dnu/4$). Right: The \'{e}chelle diagram of \object{EPIC 212708252}.    }
    \label{fig:data_ex}
\end{figure*}

Across the campaigns covered by this study, a total of $546$ observations were made in SC, spread over $492$ unique stars, resulting in the detection of solar-like oscillations in $210$ of these (see \sref{sec:seis_res}) -- in this paper we focus on the $173$ detections from C6-19. \tref{tab:fut_cam} lists the number of proposed targets and detections per campaign. We note that the overall success rate is lowered by campaigns focusing on open clusters and the inclusion of known exoplanet hosts where modest predictions for detectability were allowed.
A total of $48$ stars have been observed in two or three campaigns, and would typically have been re-proposed to improve on a positive detection of oscillations or because of an especially interesting science case. The distribution of targets largely follows that of \kp in terms of \teff and \logg (\fref{fig:kiel}), but an important distinction is that this sample peaks at a \kp magnitude ($\rm K_p$) of ${\sim}8.7$, and with very few stars having $\rm K_p>10$ (mainly M67 targets), while the sample from the nominal \kp mission peaks at $\rm K_p>11$ \citep{Mathur2021} -- making this sample more suitable for follow-up observations. \Fref{fig:sky} shows the spatial distribution of targets in the galactic frame in addition to a Toomre diagram. The targets predominantly have kinematics suggesting a thin disk origin with total velocities $V_{\rm tot}{\lesssim}70$ $\mathrm{km/s}$ \citep{Nissen2009}. Of the order ${\sim}14$ potentially belong to the thick disk with $V_{\rm tot}{\gtrsim}70$ $\mathrm{km/s}$, and $2$ with $V_{\rm tot}{\gtrsim}150$ $\mathrm{km/s}$ possibly belong to the halo (see Paper~II, Section~$5$).

During the reduction of data from K2 Cycle 4 (C11-13) we observed that for targets around a \kp magnitude of ${\sim}8$ the downloaded pixel stamp was often too small, not allowing the full flux to be captured. This realisation was communicated to the K2 team and the cause was identified as an underestimation of the \kp magnitude in the EPIC \citep{Huber2016} around this brightness, from the use of systematically incorrect APASS magnitudes (Barentsen G., private communication). A correction for this underestimation was implemented which took full effect from C17 onward. Unfortunately, the small pixel stamps resulted in an inability to detect solar-like oscillations for several bright stars in the sample, likely of the order ${\sim}50$ stars. We did try to detect oscillations using the halo photometry method \citep{pleaK2}, but this was unsuccessful.

\section{Input data} \label{sec:data}
Spectroscopic data for 163 targets (out of 173) were obtained from the Tillinghast reflector Echelle Spectrograph \citep[TRES;][]{2007RMxAC..28..129S, furesz_phd,Mink2011} on the 1.5-m Tillinghast telescope at the \textit{F.~L.~{Whipple}} Observatory on Mt. Hopkins in Arizona. TRES is a fiber-fed optical echelle spectrograph with a wavelength range $390-910$ nm and a resolving power of $R \sim 44,000$.
Astrometric data, as well as photometry for our use of the infrared flux method (IRFM; \sref{sec:irfm}) and for deriving luminosities (\sref{sec:lum}), were generally obtained from \textit{Gaia} EDR3 \citep{2016A&A...595A...1G,EDR3_2021,Riello2021}. 

Photometric data for the asteroseismic analysis (\sref{sec:ana}) were obtained from the KASOC database\footnote{\url{http://kasoc.phys.au.dk}} and light curves were made using the K2P$^2$ pipeline \citep{k2p2,2014MNRAS.445.2698H}, where the flux is decorrelated against the systematic movement across the CCD \citep{2014PASP..126..948V,2016van.cleve.pasp}. 
For most stars, we constructed custom apertures to better conserve the flux because many stars saturate the CCD causing in some cases bleeding trails. In \fref{fig:data_ex} we show an example (for \object{EPIC 212708252}) of the photometric data before and after the systematics correction, and the impact on the resulting power density spectrum used in the seismic analysis. First, this demonstrates the importance of a proper correction for the strong systematics inherent to K2 data before any asteroseismic analysis can be considered. Secondly, it shows that if properly treated, it is indeed possible to obtain high-quality data for such analysis from K2. 
We note that there are small variations in the use of quality flags in the filtering for different campaigns, mainly due to variations in the assignments from the K2 mission.
For campaigns C10 and C11, the observations were split into sub-campaigns. In the case of C10, we ended up only using data from C10.2 due to the poor quality of data in C10.1; for C11 we used all data by concatenating the sub-campaigns. For C19 we use only the last ${\sim}17$ days of data, because of the low data quality at the beginning of the campaign. For C19 we furthermore used our own calculation of flux centroids for the correction of the time series as the ones provided by the mission resulted in a poor correction for the systematic noise.


\section{Stellar parameters} \label{sec:stellarparams}

In this section, we outline the methodologies employed to determine the stellar parameters for the sample, including atmospheric parameters (\sref{sec:atmos}), luminosities (\sref{sec:lum}), and global asteroseismic quantities (\sref{sec:ana}). We detail the different techniques used to acquire these parameters, including both spectroscopic assessments and the IRFM. Emphasis is placed on evaluating systematic uncertainties and cross-validating results through comparative analyses across different methods. Each subsection presents the derived values, explores potential biases, and highlights the consistency achieved across the various methods used.

\subsection{Atmospheric parameters}\label{sec:atmos}
We obtain atmospheric parameters from both spectroscopy and the IRFM. Results from both methods are provided in \tref{tab:atmos}, and we provide a comparison in \sref{sec:compare_atmo}.

Based on the typical interval covered by our stars in $\rm [\alpha/Fe]$ from $-0.025$ to $0.05$ dex, as found from the spectroscopic surveys APOGEE, LAMOST, and GALAH (see \aref{sec:app_survey}) we generally adopt $\rm [\alpha/Fe]=0$ dex in our further analysis, hence we assume $\rm [M/H]\simeq\feh$. We include a non-zero value if $\rm [\alpha/Fe]>0.05$ dex and the corresponding \feh from the source is in agreement with our spectroscopic value (\sref{sec:spec}) -- this turns out to be the case only for \object{EPIC 228720824} (see Paper~II, Section~5, for details).

\subsubsection{Spectroscopy}\label{sec:spec}

The Stellar Parameter Classification pipeline \citep[SPC;][]{2012Natur.486..375B} was used to derive atmospheric parameters from TRES spectra (\sref{sec:data}). Several spectra were typically obtained for each star and the adopted atmospheric parameters were given by the signal-to-noise ratio (S/N) weighted average of results from individual spectra. We include also the quality factor ($QF$) recently implemented in the SPC pipeline \citep{Bieryla2024} which assigns a flag to the spectra based on a simple decision-tree taking into account the \vsini, S/N, \teff range, and the cross-correlation function (CCF). Whenever possible we include only spectra deemed ``excellent'' ($QF=1$) or ``good'' ($QF=2$). For three stars, however, we only have results from spectra deemed to be of ``fair'' ($QF=3$; \object{EPIC 212291429}) or ``poor'' quality ($QF=4$; EPICs 211409088 and 211416749) -- we caution that the SPC results for these stars should be treated with care. In the modelling (Paper~II), we use only parameters from other spectroscopic surveys and the IRFM for the $QF=4$ stars (see \sref{sec:irfm}).

\begin{figure}
    \centering
    \includegraphics[width=\columnwidth]{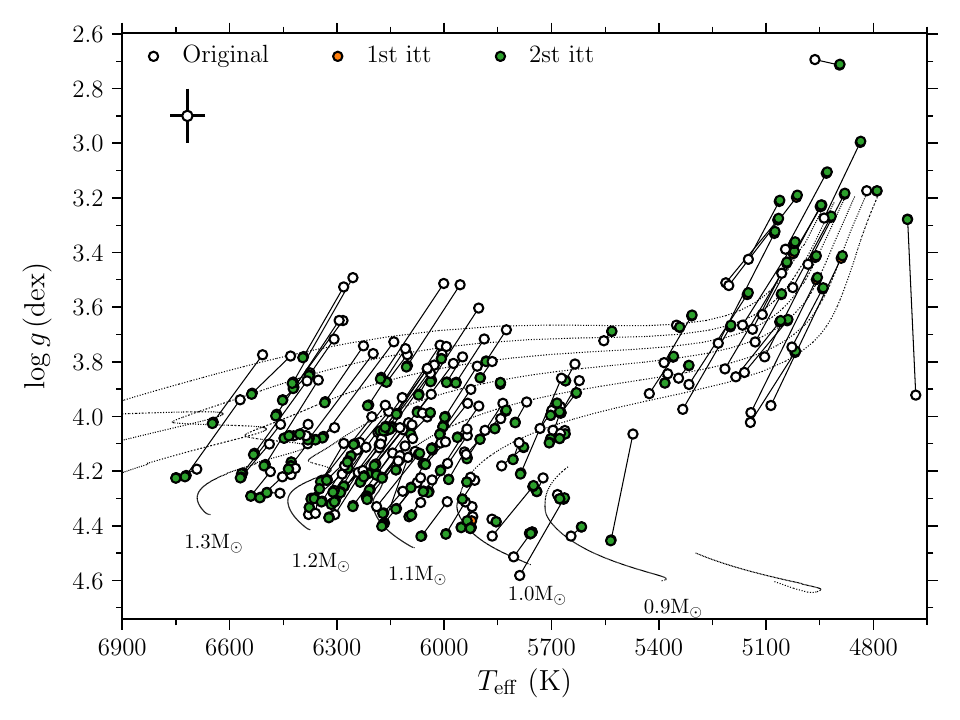}
    \caption{Change in \teff and \logg from iterating the spectroscopic reduction with a \logg fixed to the seismic values from \numax and \teff (\eqref{eq:logg}). The marker colour indicates the step of the iteration, and for each star lines connect associated values. The marker in the top left corner indicates the typical uncertainty on \logg and \teff.}
    \label{fig:spc_itthr}
\end{figure}

With the SPC analysis in hand, we proceed as in \citet{keystone} and assess the impact of iterating the spectroscopic solution with an estimate for the value of \logg based on the asteroseismic \numax, following
\begin{equation}\label{eq:logg}
g \simeq g_{\sun} \left( \frac{\numax}{\nu_{\rm max,\sun}} \right) \left( \frac{T_{\rm eff}}{T_{\rm eff, \sun}} \right)^{1/2}\, ,
\end{equation}
and using $\nu_{\rm max,\sun} = 3090\, \rm \mu Hz$, $T_{\rm eff, \sun} = 5777$ K, and $g_{\sun}=27402\, \rm cm\, s^{-2}$ \citep[][]{brown1991,kjeldsen1995,Huber2011,Chap2014}. Based on the study by \citet{Coelho2015} this relation should be accurate to within ${\sim}1.5\%$ in \numax. The reason for such an iteration is to alleviate the well-known degeneracies between spectroscopic estimates for \teff, \logg, and \feh \citep[][]{2005MSAIS...8..130S,Kordopatis2011,Torres2012}.
We iterated the SPC analysis with \logg fixed to the seismic value twice, finding that for a potential third iteration, the change in \logg would be at the level of $\pm 0.0005$ dex. In this iterative setup, only the central values for the parameters are used. Therefore, the uncertainty on \numax is not propagated to the final spectroscopic parameters. We note that while the average change in the parameters is small for the ensemble, the absolute changes range from $\pm 200$ K in \teff, $\pm 0.6$ dex in \logg, and $\pm 0.15$ dex in \feh. In \fref{fig:spc_itthr} we show the change in \teff and \logg in a Kiel-diagram. As seen, the difference between the 1st and 2nd iterations is small and difficult to discern in the plot -- already from the first to the second iteration the level of change was at $\pm 5$ K in \teff, $\pm 0.002$ dex in \feh, $\pm 0.01$ dex in \logg, and $\pm 0.01 \rm km/s$ in \vsini (see \fref{fig:spc_itt2}). The changes are generally unidirectional but with different signs when considering stars of different evolutionary stages, where MS/SG stars generally become hotter and denser and vice versa for more evolved red giants. An expected dominant source of the change at higher temperatures is given by the sensitivity of the SPC method to the pressure broadened Mg~I~b triplet near ${\sim}5200\AA$, which has weakened wings at \teff$\gtrsim6000$ K and therefore loses its sensitivity to \logg \citep{Torres2012,Brewer2015}. We refer to \aref{sec:spclogg} for further details.

The internal uncertainties from SPC are subject to error floors of $50$ K for \teff, $0.1$ dex for \logg, $0.08$ dex for \feh, and $0.5$ km/s for \vsini. These are adopted to match the expected systematic uncertainties from the specific spectroscopic analysis procedures used in SPC.
Based on the analysis of \citet{Torres2012} on the agreement between different spectroscopic analysis procedures, additional systematic uncertainties of 59 K and 0.062 dex were added in quadrature to the \teff and \feh estimates from SPC, resulting in median uncertainties of $77$ K on \teff and $0.1$ dex on \feh. We note that these uncertainties fully cover the scatter we get from different spectroscopic observations of the same star, with varying S/N. For the 41 cases where multiple spectra (between 2 and 15) were taken for a given star, we obtain standardised median absolute deviations (MADs) of the differences between individual observations and the S/N-weighted averages of ${\sim}8.4$ K in \teff, ${\sim}0.01$ dex in \feh, and ${\sim}0.12$ km/s in \vsini\ -- these values are fully in line with \citet{Brewer2018} considering the typical S/N of ${\sim}73\pm15$ for the spectra of these stars.

As a consistency check of the SPC results, we compared the extracted radial velocities (RVs) to those from \textit{Gaia} DR2 \citep{Soubiran2018} (as also adopted in \textit{Gaia} EDR3), finding an excellent agreement. We refer to \aref{sec:spcrv} for more on this comparison and here we also discuss the size of the Doppler shift on measured frequency parameters (in this paper \numax) imposed by the stellar line-of-sight velocity \citep{Davies2014}.


\subsubsection{Infrared flux method (IRFM)}\label{sec:irfm}
As an independent measure of the \teff we use the IRFM \citep{Casagrande2010,2014ApJ...787..110C,Casagrande2021} based on \textit{Gaia} EDR3 \citep{2016A&A...595A...1G,EDR3_2021,Riello2021} and $JHK_s$ photometry from the Two Micron All Sky Survey \citep[2MASS;][]{2003yCat.2246....0C,2006AJ....131.1163S}.
Here we used the official cross-match of EDR3 with 2MASS provided in \texttt{gaiaedr3.tmass$\_$psc$\_$xsc$\_$best$\_$neighbour} \citep{Marrese2021}. In applying the IRFM, we also obtain the stellar angular diameter $\theta$. Utilising $\theta$ along with a measured distance facilitates an independent calculation of the stellar radius, providing a consistency check against asteroseismically derived radii.

Reddening values were included based on the 3D dust maps of the \textit{Stilism}\footnote{\url{https://stilism.obspm.fr/}} (STructuring by Inversion the Local Interstellar Medium) project \citep{Lallement2014,Capitanio2017,Lallement2019}. We used \textit{Gaia} EDR3 distances from \citet{BJ2021}, except for five cases where \textit{Gaia} EDR3 was unavailable we had to use either \textit{Gaia} DR2 \citep{2018A&A...616A...1G} or \textsc{Hipparcos} \citep{2007A&A...474..653V} parallaxes to assess the distance, see \tref{tab:astro}. 
While the \textit{Stilism} values are typically non-zero even in the close proximity of the Sun, we set all reddening values to zero if the star is closer than $100$ pc (see \fref{fig:sky}). For stars belonging to the M67 open cluster, we used the reddening of $E(B-V)=41\pm 4$ mmag from \citet{Taylor2007}. We refer to \aref{sec:app_red} for more discussions on the reddening values tested, including a comparison with values from \texttt{Bayestar19} map \citep{Green2019}.

\setlength\tabcolsep{0pt}
\begin{table*} 
\centering 
\caption{Identifiers and astrometric parameters for the KEYSTONE sample.} 
\label{tab:astro}
\small
\begin{tabular*}{\linewidth}{@{\extracolsep{\fill}}lccclccccc@{}} 
\toprule 
\multicolumn{2}{c}{K2} & & \multicolumn{5}{c}{Gaia}  & & \\ 
\cmidrule(r){1-2}\cmidrule(l){4-9} \\ [-1em] 
EPIC &  Kp  & HIP ID & EDR3 ID & Dist & RUWE & $E(B-V)$ & $L_{\rm SPC}$ & $L_{\rm IRFM}$ & Notes \\ 
  &  (mag)  &   &   & (pc) & & (mmag) & ($\mathrm{L_{\odot}}$) & ($\mathrm{L_{\odot}}$) & \\ 
\midrule 
$201623069$ & $8.50$ & $54281$ & $3811534951212403072$ & $130.2$\rlap{$_{-0.5}^{+0.6}$} & 1.28 & $10 \pm 16$ & $5.43 \pm 0.09$ & $5.38 \pm 0.08$ & \\  [0.5ex] 
$201644284$ & $8.20$ & $60264$ & $3701419896778537216$ & $90.2 \pm 0.2$ & 0.894 &   & $3.18 \pm 0.05$ & $3.22 \pm 0.03$ & \\  [0.5ex] 
$201725213$ & $10.17$ & $54262$ & $3814827954178529664$ & $238.5$\rlap{$_{-1.0}^{+0.8}$} & 1.203 & $23 \pm 25$ & $4.20 \pm 0.10$ & $4.22 \pm 0.09$ & \\  [0.5ex] 
$203530127$ & $7.11$ & $82708$ & $6034500386022688896$ & $67.6 \pm 0.1$ & 0.882 &   & $4.72 \pm 0.05$ & $4.73 \pm 0.04$ & \\  [0.5ex] 
$211311380$ & $9.13$ & $41378$ & $600698184764497664$ & $105.6 \pm 0.2$ & 0.982 & $4 \pm 15$ & $2.44 \pm 0.03$ & $2.42 \pm 0.02$ & K2-93, S(1,2,3)\\  [0.5ex] 
$211388537$\tablefootmark{a} & $12.29$ & $$ & $604703052788343296$ & $837.7$\rlap{$_{-12.7}^{+13.0}$} & 0.909 & $41 \pm 4$\tablefootmark{b} & $8.68 \pm 0.40$ & $8.62 \pm 0.41$ & S(4)\\  [0.5ex] 
$211401787$ & $9.71$ & $$ & $601159910928534144$ & $157.9$\rlap{$_{-0.5}^{+0.4}$} & 1.059 & $9 \pm 16$ & $2.95 \pm 0.04$ & $2.90 \pm 0.04$ & \\  [0.5ex] 
$211403248$\tablefootmark{a} & $12.31$ & $$ & $604901823875341056$ & $811.4$\rlap{$_{-9.9}^{+12.0}$} & 1.023 & $41 \pm 4$\tablefootmark{b} & $8.04 \pm 0.37$ & $7.96 \pm 0.35$ & S(4)\\  [0.5ex] 
$211405262$\tablefootmark{a} & $12.66$ & $$ & $604912647193030016$ & $821.0$\rlap{$_{-10.6}^{+8.4}$} & 1.012 & $41 \pm 4$\tablefootmark{b} & $6.43 \pm 0.28$ & $6.49 \pm 0.27$ & \\  [0.5ex] 
$211409088$\tablefootmark{a} & $12.82$ & $$ & $604916770361557504$ & $836.5$\rlap{$_{-10.4}^{+12.1}$} & 1.015 & $41 \pm 4$\tablefootmark{b} & $5.38 \pm 0.24$ & $5.41 \pm 0.23$ & S(4)\\  [0.5ex] 
\bottomrule
\end{tabular*} 
\tablefoot{\small Table~\ref{tab:astro} is published in its entirety in the machine-readable format at the CDS via anonymous ftp to \url{cdsarc.u-strasbg.fr} ($130.79.128.5$) or via \url{http://cdsweb.u-strasbg.fr/cgi-bin/qcat?J/A+A/}. A portion is shown here for guidance regarding its form and content.\\
The table provides identifiers and astrometric parameters for the $173$ targets under study, sorted by EPIC ID. ``Kp'' gives the \kp magnitude \citep[][]{2011AJ....142..112B,2016ApJS..224....2H};
	``HIP ID'' and ``EDR3 ID'' give the \textit{Hipparcos} \citep{1997A&A...323L..49P} and \textit{Gaia} EDR3 \citep{EDR3_2021} identifiers of the target; ``Dist'' gives the
	photogeometric distance from \citet{BJ2021}, unless otherwise stated; ``RUWE'' gives the renormalised unit weight error from \textit{Gaia} \citep{ruwe2018}; $E(B-V)$
	gives the reddening from the 3D dust maps for the \textit{Stilism} project \citep{Capitanio2017}, unless otherwise stated. ``$L_{\rm SPC}$'' and ``$L_{\rm IRFM}$'' refer to luminosities calulated using
	\textit{Gaia} EDR3 data combined with \teff and \logg from either SPC or IRFM (see \sref{sec:lum}). In the Notes column ``RVEH'' is short for radial velocity exoplanet host; ``WDS'' is short for Washington double star;
	``S'' refers to a seismic investigation; the number in parenthesis refers to the reference listed in the table references listed below.
    \\
\tablefoottext{a}{Member of M67}; \tablefoottext{b}{M67 reddening from \citet{Taylor2007}}; \tablefoottext{c}{Distance from \textit{Gaia} DR2 \citep{BJ2018}}}
\tablebib{(1) \citet{K293van}; (2) \citet{k293}; (3) \citet{Bryant2021_k293}; (4) \citet{Stello2016};   (5) \citet{Giguere2015}; (6) \citet{HD75784-2018}; (7) \citet{Grunblatt2019}; (8) Washington Double Star Catalog \citep{WDS2001};   (9) \citet{Ong2021B}; (10) \citet{Pope2016}; (11) \citet{Kruse2019}; (12) \citet{Robinson2007}; (13) \citet{North2017};   (14) \citet{Pourbaix2004}; (15) \citet{Griffin2013}; (16) \citet{Johnson2011}; (17) \citet{Luhn2019};   (18) \citet{Tamuz2008}; (19) \citet{Moutou2011}; (20) \citet{Ginski2016}; (21) \citet{Eylen2018};   (22) \citet{Jones2021}
}
\end{table*}

Similar to the approach taken in \citet{keystone}, \teff and $\theta$ were estimated for a range of \logg values ($1 \leq \logg \leq 5$ in steps of $0.5$ dex) and metallicities ($-0.8 \leq \feh \leq 0.2$ in steps of $0.1$ dex). We note that the main sensitivity of the \teff is to \logg, and only mildly to \feh. Each point in the grid in \logg and \feh has an associated value for \teff (and $\theta$) and an uncertainty given by the scatter in \teff from the different 2MASS photometric bands, and to this we fit a 2D second-order polynomial function to describe the \teff-\logg-\feh dependence. This fit is done using \texttt{PyMC3} \citep{pymc3} with the model sampled using the No-U-Turn Sampler \citep[NUTS;][]{NUTS} assuming normally distributed errors on all coefficients of the plane.
We then sample from the coefficient of the plane in a Monte Carlo manner and in the process iterate the value of \teff by calculating a value for \logg using \eqref{eq:logg} and sampling \feh from the spectroscopic value. After only a few iterations the solution converges and the end results are distributions of self-consistent values of \teff and \logg (given the \feh from spectroscopy) from which we adopt the median and use the $68.3\%$ highest probability density (HPD) interval for the uncertainty.
The same procedure is followed for the estimation of the angular diameter $\theta$ from the IRFM, however, here we omit the dependence on metallicity.

We further add a systematic uncertainty from a Monte Carlo sampling including photometric and reddening errors. For the reddening a $20\%$ error or a Gaussian centred at $0.01$ mag was adopted, depending on which is the largest (if reddening was 0 mag from the Stilism extinction map and/or the star is closer than $100\, \rm pc$, the reddening was kept to 0 mag, but if 0 mag and further away than $100\, \rm pc$ a Gaussian centred at $0.01$ mag was adopted). We finally add zero-point uncertainties of 20 K in \teff and $0.7\%$ on $\theta$ \citep{Casagrande2010}. Combined, this results in median uncertainties of $41$ K on \teff and $2 \, \mu$as on $\theta$.

For the ten stars without a metallicity constraint from our spectroscopic analysis, we searched the literature and found metallicities for six of these. Based on a comparison with some of the large spectroscopic surveys (see \sref{sec:compare_atmo} and \aref{sec:app_survey}) we mainly used results from APOGEE DR16 \citep{apogee2020} and made a S/N-weighted average of the metallicity when multiple measurements were available. For the remaining four stars we adopted a metallicity of $\feh = -0.05 \pm 0.22$ dex, which is consistent with the metallicity distribution function of the local solar neighbourhood \citep[see, \eg,][]{2011A&A...530A.138C,2015ApJ...808..132H}. Indeed, all four stars are within ${\sim}114$ pc of the Sun, and with total galactic velocities below ${\sim}42\,\,\rm km/s$ indicating that they belong to the local solar neighbourhood\footnote{Only for \object{EPIC 226083290} was it not possible to obtain a galactic velocity from a lack of a radial velocity measurement}. We note that these ten stars are not processed in the asteroseismic analysis adopting the spectroscopic values, but the metallicities found from the literature are used to derive an IRFM \teff (\sref{sec:irfm}). The source of the atmospheric parameters is indicated in \tref{tab:atmos} if not provided by the SPC analysis.

For the three stars with $QF>2$ (EPICs $212291429$, $211409088$, and $211416749$) we also compared the results from SPC to those from the spectroscopic surveys (see \sref{sec:compare_atmo} and \aref{sec:app_survey}). For the two $QF=4$ stars (EPICs $211409088$ and $211416749$) we find significant disagreement between SPC \teff and \feh and the corresponding values from the surveys, while the surveys are in agreement with each other. Therefore, we adopt the APOGEE DR16 \citep{apogee2020} results for these stars in the IRFM calculation.
For the $QF=3$ star (\object{EPIC 212291429}) we only have external values from the Geneva-Copenhagen survey (GCS) \citep[][]{2011A&A...530A.138C}, and here find a reasonable agreement to our SPC results which therefore are kept for the IRFM analysis.

Finally, we note that for EPICs $248514180$ and $228720824$ no proper match could be made between the \textit{Gaia} EDR3 identifier and 2MASS; for \object{EPIC 249620304} the corrected version\footnote{\url{https://github.com/agabrown/gaiaedr3-flux-excess-correction}} of the \texttt{phot$\_$bp$\_$rp$\_$excess$\_$factor} is, at $0.261$, outside the recommended range $-0.08 < C^*< 0.2$ to trust \textit{Gaia} photometry \citep{Riello2021}, and for \object{EPIC 212819198} the 2MASS $H$- and $K_s$-band magnitudes are labelled as upper limits and without uncertainties. Except for $249620304$, which turned out to have an IRFM \teff in agreement with SPC, we omitted these stars from the IRFM analysis. 

\setlength\tabcolsep{0pt}
\begin{table*} 
\centering 
\caption{Atmospheric parameters for the KEYSTONE sample.
    } 
\label{tab:atmos}
\small
\begin{tabular*}{\linewidth}{@{\extracolsep{\fill}}lccccccccc@{}} 
\toprule 
\multicolumn{3}{c}{K2} & \multicolumn{2}{c}{IRFM} & \multicolumn{5}{c}{SPC} \\ 
\cmidrule(r){1-3}\cmidrule(lr){4-5}\cmidrule(l){6-10} \\ [-1em] 
EPIC & Cam. & Kp    &  $\theta$ & $\teff$ & $\teff$ &  $\log g$         &   $\feh$                  & $v\sin\, i_{\star}$ & LOS  \\ 
     &      & (mag) &  ($\mu$as)    & (K)           & (K)  & ($\rm cgs$; dex)  & (dex) & ($\rm km\ s^{-1}$) & ($\rm km\ s^{-1}$) \\ 
\midrule 
$201623069$ & $14$ & $8.50$ & $151 \pm 3$ & $6029 \pm 40$ & $5827 \pm 77$ & $3.98 \pm 0.10$ & $0.06 \pm 0.10$ & $7.07 \pm 0.50$ & $7.47 \pm 0.07$\\  [0.5ex] 
$201644284$ & $10$ & $8.20$ & $217 \pm 3$ & $5329 \pm 30$ & $5383 \pm 77$ & $3.88 \pm 0.10$ & $0.19 \pm 0.10$ & $2.82 \pm 0.50$ & $-16.24 \pm 0.01$\\  [0.5ex] 
$201725213$ & $14$ & $10.17$ & $93 \pm 2$ & $5278 \pm 37$ & $5359 \pm 78$ & $3.78 \pm 0.10$ & $0.07 \pm 0.10$ & $3.61 \pm 0.50$ & $13.88 \pm 0.08$\\  [0.5ex] 
$203530127$ & $11$ & $7.11$ & $259 \pm 4$ & $6212 \pm 45$ & $6180 \pm 77$ & $4.05 \pm 0.10$ & $0.15 \pm 0.10$ & $8.59 \pm 0.50$ & $-22.22 \pm 0.05$\\  [0.5ex] 
$211311380$ & $18$ & $9.13$ & $114 \pm 3$ & $6339 \pm 50$ & $6307 \pm 77$ & $4.31 \pm 0.10$ & $-0.04 \pm 0.10$ & $7.07 \pm 0.50$ & $50.70 \pm 0.04$\\  [0.5ex] 
$211388537$\tablefootmark{a} & $18$ & $12.29$ & $41 \pm 1$ & $5046 \pm 41$ & $5021 \pm 77$ & $3.40 \pm 0.10$ & $-0.06 \pm 0.10$ & $3.19 \pm 0.50$ & $34.05 \pm 0.07$\\  [0.5ex] 
$211401787$ & $18$ & $9.71$ & $83 \pm 2$ & $6336 \pm 53$ & $6213 \pm 77$ & $4.21 \pm 0.10$ & $-0.09 \pm 0.10$ & $9.26 \pm 0.50$ & $10.77 \pm 0.02$\\  [0.5ex] 
$211403248$\tablefootmark{a} & $16,18,A$ & $12.31$ & $41 \pm 2$ & $5023 \pm 38$ & $5042 \pm 77$ & $3.44 \pm 0.10$ & $0.01 \pm 0.10$ & $3.29 \pm 0.50$ & $33.52 \pm 0.10$\\  [0.5ex] 
$211405262$\tablefootmark{a} & $\textit{16},18,A$ & $12.66$ & $35 \pm 2$ & $5091 \pm 39$ & $5150 \pm 82$ & $3.55 \pm 0.10$ & $0.14 \pm 0.10$ & $3.47 \pm 0.50$ & $34.18 \pm 0.10$\\  [0.5ex] 
$211409088$\tablefootmark{a,*} & $\textit{5},\textit{16},18,A$ & $12.82$ &  &  & $5198 \pm 102$ & $3.67 \pm 0.14$ & $0.27 \pm 0.10$ & $5.31 \pm 0.52$ & $33.20 \pm 0.15$\\  [0.5ex] 
 -- & {} &  & $31 \pm 2$ & $5136 \pm 36$ & $5160 \pm 100$\tablefootmark{b} & $3.72 \pm 0.07$\tablefootmark{b} & $-0.01 \pm 0.01$\tablefootmark{b} &  & \\  [0.5ex] 
\bottomrule
\end{tabular*} 
\tablefoot{Table~\ref{tab:atmos} is published in its entirety in the machine-readable format at the CDS via anonymous ftp to \url{cdsarc.u-strasbg.fr} ($130.79.128.5$) or via \url{http://cdsweb.u-strasbg.fr/cgi-bin/qcat?J/A+A/}. A portion is shown here for guidance regarding its form and content.\\
The table provides atmospheric parameters for the $173$ targets under study, sorted by EPIC ID. ``Cam'' gives the K2 campaign(s) during which a given target was observed in SC, and indicates which individual campaigns resulted in a detection of oscillations -- no detections were made from individual campaigns written in italics. An ``A'' indicates that the seismic values are based on the combined data from all campaigns. We note that analysis was not performed for individual	C5 data, but this data was included if available when combining all data for a given star. ``$\theta$'' gives the stellar angular diameter from the IRFM in $\mu$as. Results from the IRFM are generally based on especially SPC \feh and $E(B-V)$ as listed in \tref{tab:astro};	``LOS'' gives the line-of-sight velocity from the CfA TRES observations, corrected by $-0.61\, \rm km/s$. Deviations from the standard sources mentioned here are stated in the table, see the footnotes. A repeated entry for a given star (in which the EPIC is not indicated again) gives the spectroscopic values adopted from a literature study, and the associated IRFM results, in cases where the quality of the SPC data was deemed too low.\\
\tablefoottext{a}{Member of M67}; \tablefoottext{b}{Parameters from APOGEE DR16 \citep{apogee2020}}; \tablefoottext{c}{Parameters from \citet{Luck2017}, uncertainty on \feh assigned based on spread from GCS \citep{2011A&A...530A.138C}}; \tablefoottext{d}{GCS average metallicity \citep{2011A&A...530A.138C}}; \tablefoottext{e}{RV from \textit{Gaia} DR2 \citep{Soubiran2018}}; \tablefoottext{f}{poor 2MASS-EDR3 cross-match}; \tablefoottext{g}{high \texttt{phot$\_$bp$\_$rp$\_$excess$\_$factor} on \textit{Gaia} photometry}; \tablefoottext{h}{poor 2MASS photometry}; \tablefoottext{*}{SPC has poor quality factor}
}
\end{table*}


\subsubsection{Comparison of input atmospheric parameters}\label{sec:compare_atmo}
\begin{figure*}
    \centering
    \includegraphics[width=1\textwidth]{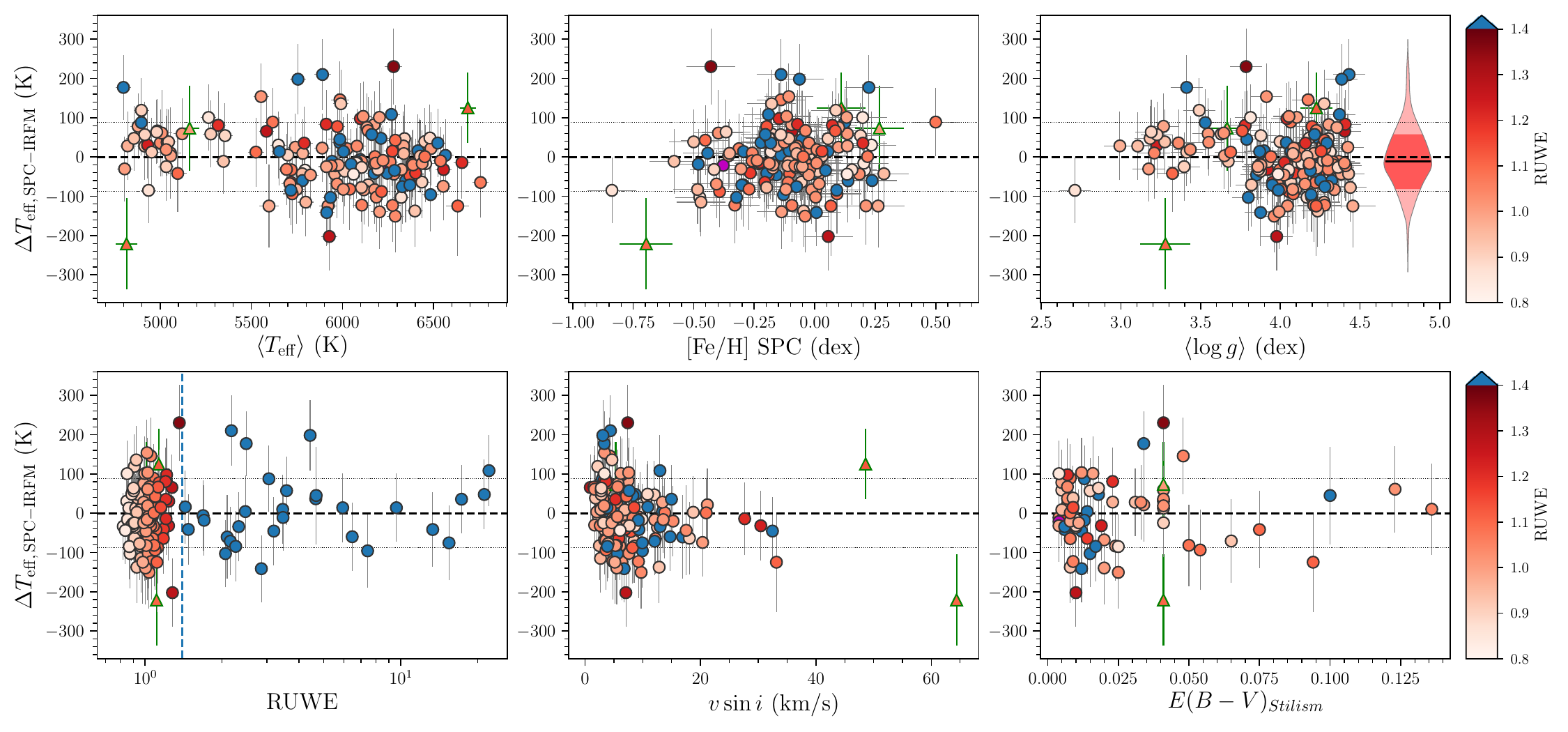}
    \caption{Comparison of \teff values obtained from our spectroscopic analysis (\sref{sec:spec}) and from the IRFM (\sref{sec:irfm}), plotted against the different atmospheric parameters and the external parameters RUWE and $E(B-V)$ (that could impact especially the IRFM). The differences are coloured by the RUWE value from \textit{Gaia} EDR3 \citep{ruwe2018}, with the colour capped at a RUWE value of $1.4$ (marked in the bottom left panel as a vertical dashed line), which is suggested as the upper limit for a single source with a non-problematic astrometric solution For \object{EPIC 231478973} no RUWE value is available, and this star has been indicated by a magenta marker. The horizontal dashed line marks the zero-difference, while the horizontal dotted lines mark the median \teff uncertainty of the differences. The violin insert in the upper right panel shows the distribution of the differences, with the darker red interval indicating the standardized MAD interval. The triangular markers with green errorbars mark the three stars with likely unreliable SPC results.}
    \label{fig:teff_compare}
\end{figure*}
For the 160 stars with both SPC and IRFM results \fref{fig:teff_compare} provides a comparison of the \teff values. To enable better visual identification of potential proportional biases \citep{bland1986statistical} we plot the differences in \teff against the average \teff and \logg values, and against the SPC \feh values. 
There is an overall excellent agreement between the \teff estimates. The median difference for the sample is only $-11$ K (IRFM \teff being higher than the spectroscopic ones) and the standardised MAD of the differences is $65$ K, which should be compared to the median uncertainty on the differences of $88$ K.

From \fref{fig:teff_compare} there appears to be a proportional bias between the \teff estimates, where positive differences are over-represented at $\langle \teff \rangle$ smaller than approximately the solar \teff, and vice versa for larger $\langle \teff \rangle$. To quantify the relation between the two \teff estimates we applied a Bayesian errors-in-variables regression analysis (see Paper~II, Section $4.2.1$). From this analysis we found a relation as $T_{\rm eff,\, IRFM} \approx 1.035\times T_{\rm eff,\, SPC} - 203$ K, which at the limits of the \teff interval for our sample amounts to absolute differences of ${\sim}\pm35$ K. This small, but significant, bias is mainly driven by the cool evolved stars -- if we focus the analysis to the MS/SG stars (excluding stars having  $\teff<5500$ and $\logg<4$) we obtain $T_{\rm eff,\, IRFM} \approx 1.008\times T_{\rm eff,\, SPC} - 35$ K, which amounts to differences between $9-18$ K in the interval from $5500-6700$ K. The trends seen in the relation between the IRFM and spectroscopic \teff scales are similar to those found by \citet{Sahlholdt2018} \citep[see also][]{Huber2017} but with a better overall agreement and reduced scatter from the anchoring the spectroscopic analysis to the seismic \logg (see \aref{sec:spclogg}).

We test also for proportional biases in the \teff differences against other parameters using Spearman's rank correlation $\rho$ \citep{spearman}, which quantifies the degree to which the ranked variables are monotonically associated. For this analysis, we omit the \teff differences from the two $QF=4$ stars with suspected unreliable SPC results. For $\feh$, and $\langle \logg \rangle$ as shown in \fref{fig:teff_compare}, we fail to reject the null hypothesis (H0), which assumes that the parameters are uncorrelated at the $5\%$ level ($95\%$ confidence). 

\begin{figure}
    \centering
    \includegraphics[width=1\columnwidth]{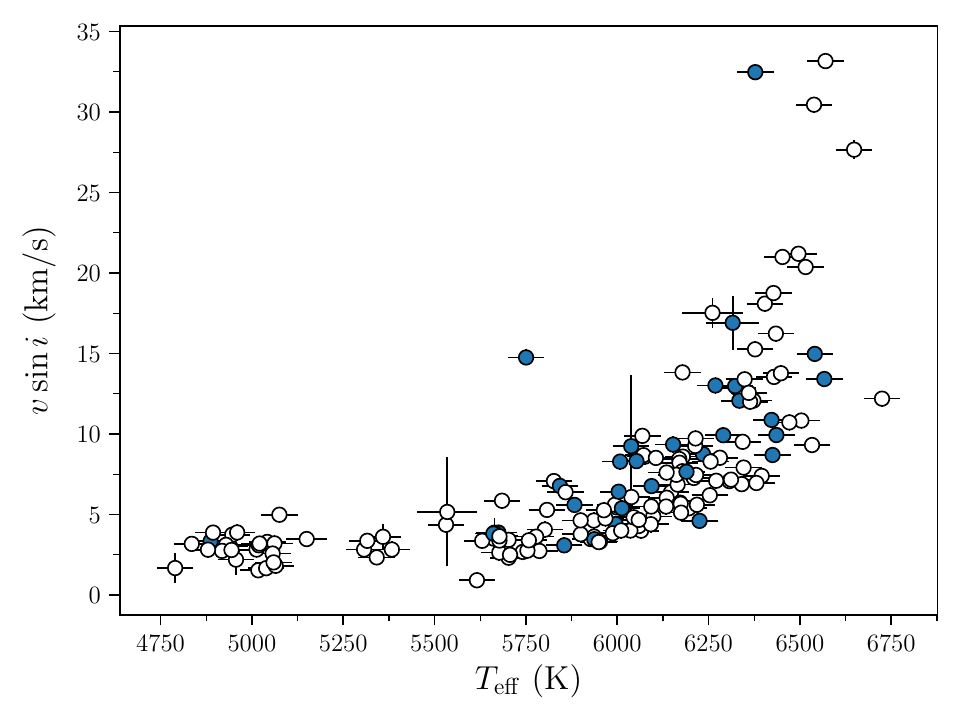}
    \caption{Relation between \teff and \vsini from the \logg-iterated SPC analysis. Here we have included only the stars with $QF=1, \, 2$. Stars with filled blue markers have a \textit{Gaia} EDR3 $\rm RUWE>1.4$. For stars with multiple observations, we have for the uncertainty on \vsini added in quadrature the standardised MAD from individual observations relative to the S/N-weighted average.}
    \label{fig:vsini}
\end{figure}

We also tested for biases with (1) the SPC $\vsini$ values, which in particular can impact the spectroscopic analysis; (2) the \textit{Gaia} RUWE (renormalised unit weight error) parameter \citep{ruwe2018} (\tref{tab:astro}), which is a good tool for identifying possible binary companions \citep{Bel2020} -- in turn, a cool companion star could add an excess infrared flux, hence affecting the IRFM; and (3) the reddening $E(B-V)$, which is known to affect the IRFM at the level of increasing \teff by ${\sim}50$ K for a $0.01$ mag increase in the reddening. Only for $\vsini$ do we see a (negative) correlation that allows H0 to be rejected at the $5\%$ level. It is, however, not surprising to see a similar proportional bias for $\vsini$ and $\langle \teff \rangle$ given the strong evolutionary (positive rank) correlation between these parameters ($\rho{\sim}0.9$) for a given stellar age, as shown in \fref{fig:vsini}. We note that when adopting the reddening values from the \citet{Green2019} \texttt{Baystar19} map, rather than the Stilism values, we also see a significant negative correlation against $E(B-V)$.

In \aref{sec:app_survey} we provide a comparison between the spectroscopic values from our analysis to those from the larger spectroscopic surveys that overlap with our sample, noting here that within uncertainties our \teff, \logg, and \feh values agree with both APOGEE DR16 \citep{apogee2020}, the GCS \citep[][]{2011A&A...530A.138C}, LAMOST \citep{lamostK22020}, and GALAH \citep{galah2021}.

\subsection{Luminosities} \label{sec:lum}

As an additional constraint we derive luminosities with \textit{Gaia} EDR3 $G$-band as our primary source of photometry. Following \citet{Torres2010} the absolute magnitude is given as
\begin{equation}
    M_{G} = -2.5 \log_{10}\left(\frac{L}{L_{\odot}}\right) + V_{\odot} + 31.572 - BC_G + BC_{V, \odot}\, .
\end{equation}
By rewriting the absolute $G$-band magnitude ($M_G$) in terms of the distance modulus we can write the luminosity as:
\begin{equation}\label{eq:lum}
    L/L_{\odot} = 10^{0.4\left(5\log_{10}(d) - G + A_G - BC_G + V_{\odot} + 26.572 + BC_{V, \odot}\right)}\, ,
\end{equation}
where $d$ is the distance in pc, $G$ is the apparent \textit{Gaia} EDR3 $G$-band magnitude, $A_G$ is the extinction in the $G$-band, and $BC_G$ is the bolometric correction. With a few exceptions noted in \tref{tab:astro}, we use photogeometric distances from \citet{BJ2021}, which incorporate the recommended parallax zero-point corrections of \citet{Lindegren2021}. We adopt values of $V_{\odot}=-26.74\pm 0.01$ mag and $BC_{V, \odot}=-0.078\pm0.005$ mag from analysis of empirical solar spectra (see \aref{app:lum} for details). 

We make saturation corrections to the \textit{Gaia} photometry following \citet{Riello2021}, and also checked the need for corrections to stars with 2- or 6-parameter solutions (corresponding to \texttt{astrometric$\_$params$\_$solved} values of 3 or 95, see \citet{EDR3_2021}) -- while 8 stars have such solutions they are all brighter than $G=13$ mag and therefore do not require a correction.

The extinction in a given band $\xi$ is computed as $A_{\xi} = R_{\xi} E(B-V)$, where the ratio of total to selective extinction $R_{\xi}$ is found from a \teff- and \feh-dependent relation similar to \citet{Casagrande2018}, but with revised coefficients for applicability to \textit{Gaia} EDR3 (see \aref{app:lum} for details). For the reddening, we adopt an uncertainty of $20\%$ on the reddening value (see \aref{sec:app_red}). 

We adopt the $R_{\xi}$ values resulting from using the \citet{Cardelli1989} extinction law. However, to capture the impact on the choice of extinction law, we add in quadrature (to the uncertainty propagated from the uncertainties in \teff, \feh, and \logg) a systematic uncertainty given by the change in $R_{\xi}$ from assuming instead the \citet{Fitzpatrick1999} extinction law (renormalized as per \citet{Schlafly2016}). In median, this systematic term contributes a ${\sim}5\%$ increase in the uncertainty on the luminosity.

For the bolometric correction $BC_G$ we use the interpolation routines of \citet{Casagrande2018}\footnote{\url{https://github.com/casaluca/bolometric-corrections}}, and adopt $R_V=3.1$. To estimate the uncertainty on the bolometric correction we perform Monte Carlo sampling of the input parameters for the interpolation routines and adopt the distribution mean and standard deviation for $BC_G$ in \eqref{eq:lum}.

\begin{figure}
    \centering
    \includegraphics[width=\columnwidth]{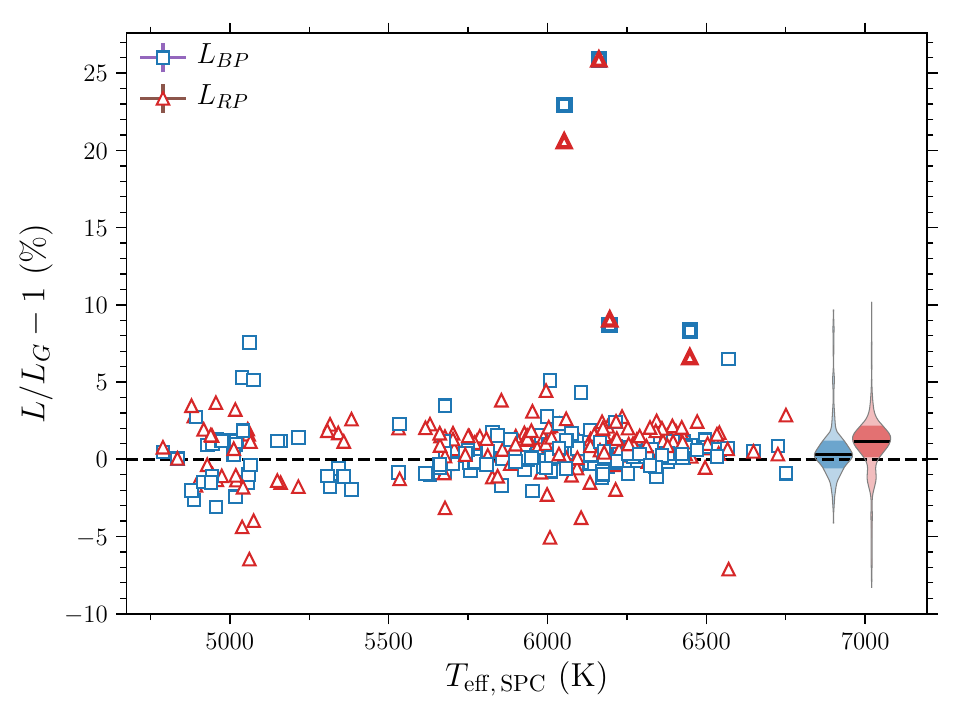}
    \caption{Relative differences as a function of \teff between luminosities calculated from the \textit{Gaia} EDR3 $BP$ and $RP$ bands relative to those from the $G$-band. Spectroscopic values from SPC were used in the analysis. The violin inserts show the distributions of the relative differences, with the median indicated by a full black line and the standardised MAD by the darker-coloured interval. Markers with thick line widths show stars with a corrected version\textsuperscript{*} of \texttt{phot$\_$bp$\_$rp$\_$excess$\_$factor}$>0.05$ \citep[see][]{Riello2021}, indicating potentially poor photometry.}
    \label{fig:lcomp}
    \noindent\tiny\textsuperscript{*} \href{https://github.com/agabrown/gaiaedr3-flux-excess-correction}{https://github.com/agabrown/gaiaedr3-flux-excess-correction}
\end{figure}

As a consistency check, we calculated also the luminosities from \textit{Gaia} EDR3 $BP$ and $RP$ bands. 
\Fref{fig:lcomp} provides a comparison between the luminosities obtained from these bands relative to those from the $G$-band. As seen the agreement is excellent, with a median relative difference of $1.15\pm1.03\%$ for the $RP$-band, and typically with the $G$- and $BP$-band values in close agreement, and with a median relative difference of $0.30\pm0.94\%$. To support our choice of the $G$-band values as our primary source of \textit{Gaia} EDR3 photometry for computing luminosities we find that $L_{G}$ has the smallest scatter as compared to the mean of the different estimates and with no apparent systematic in terms of \teff nor magnitude; the apparent $G$-band magnitudes are generally constructed from an order of magnitude more observations than the $BP$- and $RP$-band counterparts, and the photometric signal-to-noise ratio is generally factor of ${\sim}2$ higher.  

As a further consistency check, we also computed luminosities with $R_{\xi}$ values based on the \texttt{bp$\_$rp}-dependent relations by \citet{Casagrande2021}. We find in all cases full consistency between these luminosities and those using $R_{\xi}$ from the \citet{Casagrande2018} relations.	

We derive sets of luminosities using \teff and \logg from both SPC and IRFM (using in all cases \feh from SPC). The luminosities are used as an additional constraint in some versions of the modelling presented in Paper~II  (their Table~1) and are provided in \tref{tab:astro}. 

\subsection{Asteroseismic parameters} \label{sec:ana}

For the asteroseismic analysis, we focus on the global seismic parameters \dnu and \numax. 
We employ three different methods for the extraction of these parameters to identify outliers and to get a better handle on the systematic uncertainty from the choice of analysis method. The use of several independent extraction pipelines is especially important given the well-known instrumental noise of K2 data of which some residuals will typically survive into the final de-trended light curve. 

We note that several stars are of high enough quality to allow a detailed peakbagging of individual modes of oscillation (\fref{fig:data_ex}), but here we will focus on the full sample for which only the global seismic parameters can be extracted for all stars. We refer to \citet{Ong2021B} for the detailed analysis of a subset of the best stars (see \tref{tab:astro}). 

\begin{figure}[ht!]
    \centering
    \includegraphics[trim=0.5cm 0 1cm 0cm,clip,width=1\columnwidth]{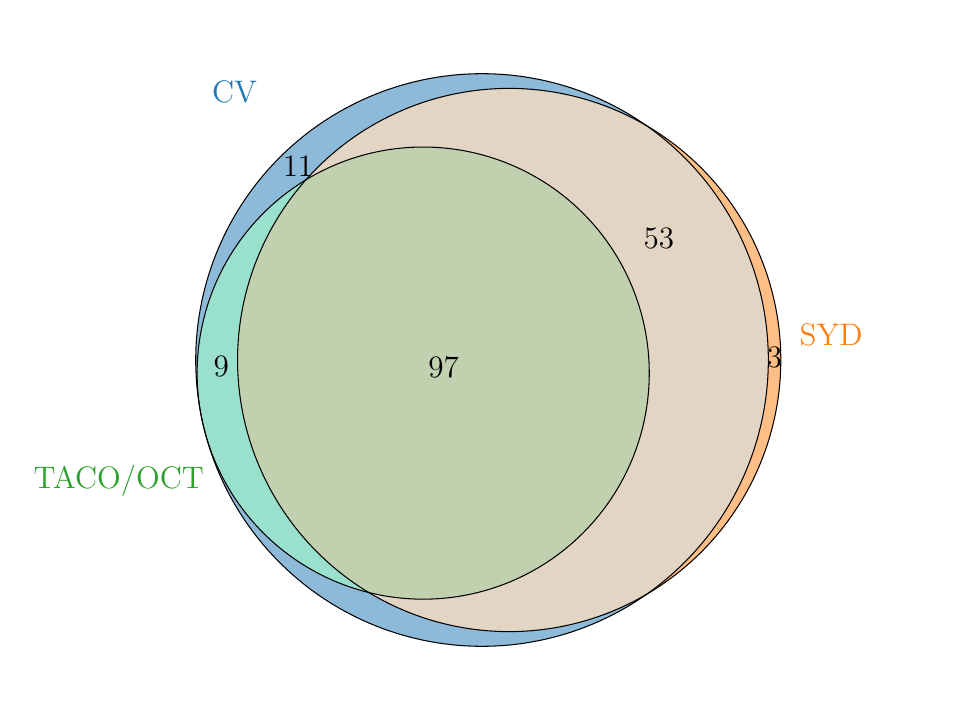}
    \caption{Venn-diagram of detections from the different analysis pipelines after a manual pruning of the claimed detections.}
    \label{fig:venn}
\end{figure}
\begin{figure*}
    \centering
    \includegraphics[width=\textwidth]{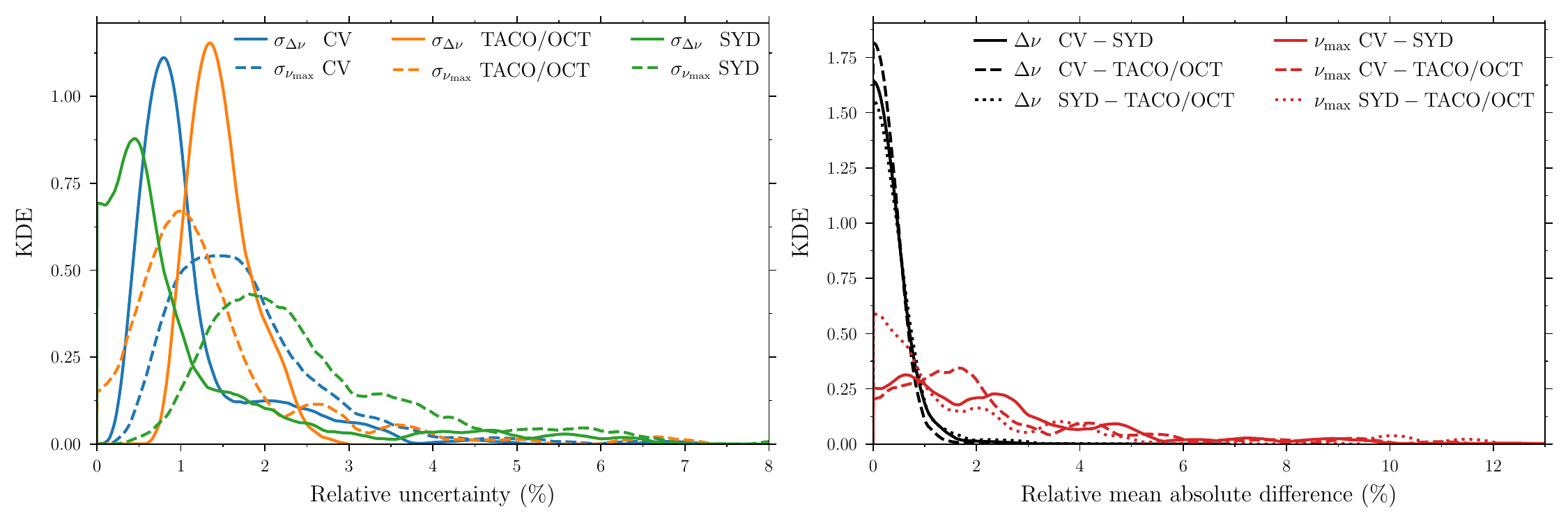}
    \caption{Comparison of uncertainties and differences for the global asteroseismic parameters. Left: Kernel density estimators (KDEs) of the relative internal uncertainty on \dnu (full lines) and \numax (dashed lines) from the different methods (see legend). Right: KDEs of the relative mean absolute differences between \dnu and \numax values from each pair of methods (see legend).}
    \label{fig:avunc}
\end{figure*}

\subsubsection{Methods}
Below we describe the three methods used to measure global seismic parameters.

\texttt{CV:} This set of seismic parameters was extracted using the coefficient of variation (CV) method \citep{Bell2019}, as implemented in \citet{viani2019}. Rather than adopting the centroid of the CV peak as our measurement for \numax, as done by \citet{viani2019}, we adopt the position of the centre of a Gaussian function fitted to the CV peak. With this modification, we obtain the uncertainty on \numax from the full width at half maximum (FWMH) and amplitude ($A$) of the Gaussian function as \citep[see][]{Garnir1987}:
\begin{equation}
    \sigma_{\numax}=0.412 \sqrt{{\rm FHWM} / A}\, .
\end{equation}
The FWHM relates to the standard deviation ($C$) of the Gaussian as ${\rm FHWM} = 2\sqrt{2\ln 2} C$. With the frequency range on oscillations identified from \numax the value of \dnu is computed from the power spectrum of the power spectrum ($\rm PS\otimes PS$) following \citet{OCTAVE}.

\texttt{SYD:} For the second set of seismic results we used the SYD pipeline \citep{2009CoAst.160...74H}. We used a frequency range between $100$ and $7000$ $\rm \mu Hz$ and modelled the granulation background with a two-component Harvey model with the white-noise component fixed to the mean value measured between $6800$ and $7000\,\,\rm \mu Hz$. We measured \numax\ as the peak of a heavily smoothed, background-corrected power spectrum and \dnu\ from the autocorrelation of the background-corrected power spectrum centred on \numax. Uncertainties on \dnu and \numax were calculated using Monte Carlo simulations as described in \citet{Huber2011}. For a subset of detections, we confirmed that the derived parameters are consistent with \texttt{pySYD} \citep{Chontos21}, an open-source python-based implementation of the SYD pipeline which uses a model-selection-based approach for fitting the granulation background. 

\texttt{TACO/OCT:}
The third set of results was derived based on a combination of the TACO (Tools for the Automated Characterisation of Oscillations; Hekker et al. in prep) and the OCT code \citep{OCTAVE}. An estimate for \numax is first searched for using several different approaches, \ie, from the variance of the flux \citep{Hekker2012}, the maxima of a Morlet, and a Mexican hat wavelet transform of the PDS. These estimates are combined for a fit of the stellar granulation background in which an MCMC algorithm is used to fit for three background components, white noise, and the oscillation power excess. To account for the possibility that the \numax estimate is off, we also fit at the position of the knees of the second and third background components and make a fit without the oscillation power excess. Based on the log-likelihood we select a best fit and use that (in case the presence of oscillation power excess is more likely than no oscillations) to select the frequency range of the oscillations. The value of \dnu is computed from the power spectrum of the power spectrum ($\rm PS\otimes PS$) following \citet{OCTAVE}
Finally, the results are inspected by eye to remove other signals that may have been picked up, such as instrumental signatures or binaries.


\subsubsection{Results}\label{sec:seis_res}

Based on observations from C6-C19, the \texttt{CV} method returned detections for 192 stars, the \texttt{SYD} method returned detections for 155 stars, while the \texttt{TACO/OCT} method returned detections for 109 stars. We note that in some cases a detection was only obtained after joining data from several campaigns (see Tables~\ref{tab:atmos} and \ref{tab:all_seis}).

In all cases of a claimed detection, we manually inspected the data for signs of excess power from oscillations, with a particular focus on the cases where only one method returned a detection. In the inspection, we visually checked the PDS, the power-of-power (PS$\otimes$PS) spectrum, and the \'{e}chelle diagram of the PDS around the claimed \numax. Additionally, we compared the claimed \numax with the predicted value from the proposal and the estimate from the spectroscopic \teff and \logg (see \sref{sec:spec}). This step is important given the systematic noise inherent to K2 data. Based on this step we discarded 25 targets and ended up with 173 stars with detections. Global seismic parameter measurements from all pipelines and all campaigns are available in \tref{tab:all_seis}.

\setlength\tabcolsep{0pt}
\begin{table*} 
\centering 
\caption{Global asteroseismic parameters for the KEYSTONE sample.} 
\label{tab:all_seis}
\begin{tabular*}{\linewidth}{@{\extracolsep{\fill}}lccccccccc@{}} 
\toprule 
\multicolumn{2}{c}{K2} & \multicolumn{2}{c}{CV} & \multicolumn{2}{c}{SYD} & \multicolumn{2}{c}{TACO/OCT} \\ 
\cline{1-2}\cline{3-4}\cline{5-6}\cline{7-8} \\ [-6pt]  
EPIC & Cam. & \numax   & \dnu     & \numax   & \dnu     & \numax   & \dnu     \\ [1pt]  
     &      & ($\muhz$) & ($\muhz$) & ($\muhz$) & ($\muhz$) & ($\muhz$) & ($\muhz$) \\ 
\midrule 
$201623069$ & 14 & $1064.0 \pm 22.8$ & $58.3 \pm 0.6$ & $1085.9 \pm 28.2$ & $58.1 \pm 0.4$ & $1061.8 \pm 18.0$ & $58.2 \pm 0.9$\\  [0.5ex] 
$201644284$ & 10 & $881.4 \pm 6.5$ & $48.6 \pm 0.9$ & $865.6 \pm 12.7$ & $48.0 \pm 0.2$ & $863.9 \pm 4.9$ & $48.4 \pm 0.7$\\  [0.5ex] 
$201725213$ & 14 & $706.2 \pm 9.9$ & $41.2 \pm 0.6$ &  &  & $712.4 \pm 8.9$ & $41.5 \pm 0.6$\\  [0.5ex] 
$203530127$ & 11 & $1231.7 \pm 19.1$ & $62.0 \pm 0.5$ & $1163.3 \pm 50.5$ & $61.9 \pm 0.3$ & $1237.4 \pm 32.9$ & $62.0 \pm 0.7$\\  [0.5ex] 
$211311380$ & 18 & $2218.2 \pm 35.6$ & $99.9 \pm 0.6$ & $2157.6 \pm 43.5$ & $100.2 \pm 0.8$ &  & \\  [0.5ex] 
$211388537$ & 18 & $300.9 \pm 7.7$ & $21.1 \pm 0.4$ &  &  & $310.3 \pm 5.7$ & $21.1 \pm 0.4$\\  [0.5ex] 
$211401787$ & 18 & $1764.4 \pm 23.0$ & $85.7 \pm 0.6$ & $1713.9 \pm 102.4$ & $85.3 \pm 0.7$ & $1748.7 \pm 30.2$ & $85.5 \pm 1.1$\\  [0.5ex] 
$211403248$ & 16 & $334.6 \pm 7.0$ & $21.5 \pm 0.5$ &  &  &  & \\  [0.5ex] 
$211403248$ & 18 & $333.8 \pm 4.2$ & $20.9 \pm 0.6$ &  &  & $319.2 \pm 4.0$ & $21.1 \pm 0.4$\\  [0.5ex] 
$211403248$ & A & $328.5 \pm 5.0$ & $21.5 \pm 0.5$ & $319.7 \pm 5.5$ & $21.5 \pm 1.1$ &  & \\  [0.5ex] 
$211405262$ & 18 & $418.3 \pm 9.8$ & $26.6 \pm 0.5$ &  &  &  & \\  [0.5ex] 
$211405262$ & A & $420.8 \pm 11.3$ & $26.5 \pm 0.5$ &  &  &  & \\  [0.5ex] 
\bottomrule
\end{tabular*} 
\tablefoot{Table~\ref{tab:all_seis} is published in its entirety in the machine-readable format at the CDS via anonymous ftp to \url{cdsarc.u-strasbg.fr} ($130.79.128.5$) or via \url{http://cdsweb.u-strasbg.fr/cgi-bin/qcat?J/A+A/}. A portion is shown here for guidance regarding its form and content. A portion is shown here for guidance regarding its form and content.\\
The table provides global asteroseismic parameters for the $173$ targets under study, sorted by EPIC ID. ``Cam'' gives the K2 campaign associated with a given measurement -- for targets with measurements from observations in multiple campaigns all measurements will be provided, hence a given star can have several entries. An ``A'' indicates the seismic values based on the combined data from all available campaigns (see \tref{tab:atmos}). We note that analysis was not performed for individual C5 data, but this data was included if available when combining all data for a given star.}
\end{table*}

\Fref{fig:venn} shows a Venn diagram of the overlaps of detections from the different methods following the manual pruning described above. From the total number of 173 detections, 97 (${\sim}56\, \%$) have results from all three pipelines, 62 (${\sim}36\, \%$) have from only two pipeline, while 14 (${\sim}8\, \%$) have results from only one pipeline. As seen, nearly all detections are found by the \texttt{CV} method, and there are no apparent systematic differences in terms of the magnitude nor \numax ranges for the detections of the different methods (not shown).  
\begin{figure}[ht!]
    \centering
\includegraphics[width=\columnwidth]{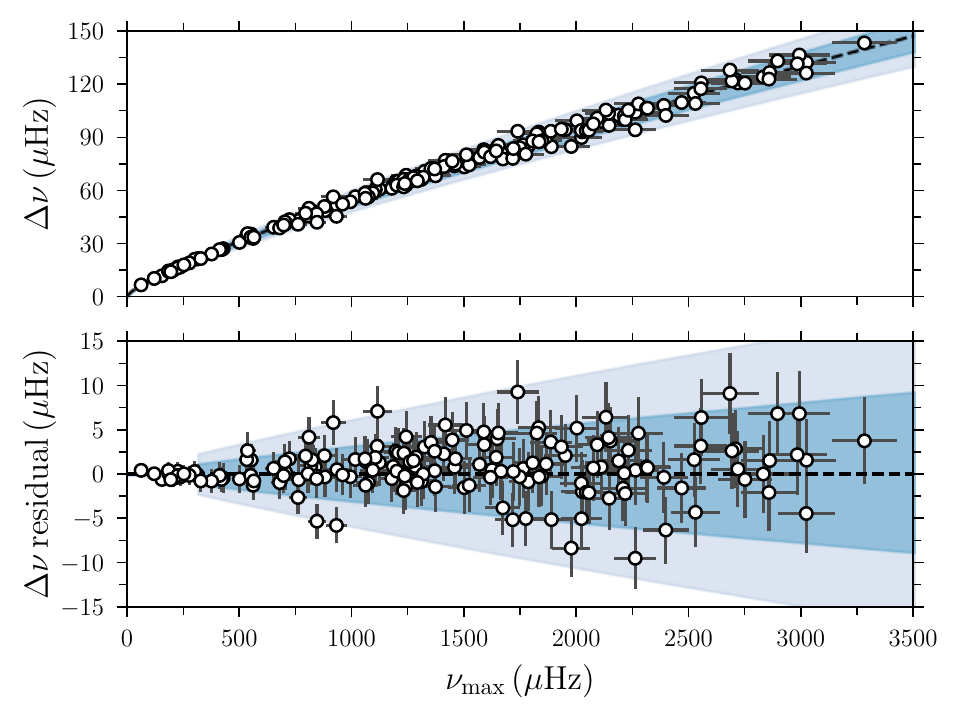}
    \caption{Correspondence between \dnu and \numax from the \texttt{CV} method (top) and residuals against the expectation from the empirical relation by \citet{Huber2011} (bottom). The light and dark coloured bands indicate the $1-$ and $2-\sigma$ uncertainties on the empirical relation. For the residuals, we have propagated the \numax uncertainty to the corresponding uncertainty on \dnu.}
    \label{fig:dnunumax}
\end{figure}

The distributions for the internal uncertainties of each method are shown in \fref{fig:avunc} (left panel). As seen the typical fractional uncertainty on \numax is of the order ${\sim}1.6\%$, and for \dnu of the order ${\sim}0.7\%$, which is in line with previous results from the literature \citep[\eg,][]{verner2011,Chap2014,serenelli2017}.

As a measure of the agreement between methods the right panel of \fref{fig:avunc} shows the distributions for the relative mean absolute differences (RMD) between the values from the different methods, weighted by the combined uncertainties of the methods, providing a normalised value for the dispersion. Given that the RMD is not based on any central tendency, any constant bias offsets between the methods will also be included in the measure of the difference. Overall we find an excellent agreement between the values from the different approaches, with maximum weighted median RMD values of ${\sim}2.2\%$ in \numax and ${\sim}0.3\%$ in \dnu in the comparisons involving the \texttt{CV} method. We also tested for a constant bias between the methods but found no significant offset.

As a measure for the systematic uncertainty from the choice of method, we computed the weighted root-mean-square (RMS) deviations between these, where the contribution from each method to the RMS is weighted by the inverse variance of the value from that method. In line with the RMD values, we obtain median relative RMS values of ${\sim}2.5\%$ in \numax and ${\sim}0.35\%$ in \dnu, where the \texttt{CV} method was used as reference. 
In general, we find a good agreement (within ${\sim}10\%$) between the estimated \numax from the different pipelines and the expected value from our target selection procedure (\sref{sec:select}), see \aref{sec:app1} for additional details.

Lastly, we test for consistency in the results returned by a single method from multiple observing campaigns. Such a comparison is shown in \fref{fig:av_multi_comp} in \aref{sec:app1} for results from the \texttt{CV} method (similar results are found from the other methods). The median difference between values from individual campaigns and multiple campaigns is consistent with zero. The scatter in the differences, based on the standardized MAD, amounts to $0.4\%$ in \dnu and $2.1\%$ in \numax, so of the same order as the scatter between methods. In this comparison, we note that the data from C19 only constitute ${\sim}17$ days of observations.

In 30 cases we have stars with detectable oscillations that have been observed over multiple campaigns. For these stars, we adopt the \dnu and \numax obtained from the weighted averaged PDS from the different campaigns, with the weights given by the inverse of the overall variance of the campaign (see \aref{sec:app1}). This version of the PDS was found to best enable the detection of oscillations, at the cost of not significantly reducing the internal uncertainties on the measured parameters.

\Fref{fig:dnunumax} shows the relationship between \dnu and \numax values from the \texttt{CV} method, together with the expectation given by the empirical relation by \citet{Huber2011}, which is fully met. The remaining scatter seen in the residuals is mainly caused by the residual dependence on mass, \teff, and luminosity in the relation between \dnu and \numax.

\section{Conclusion} \label{sec:con}

With this first set of results from the KEYSTONE project we deliver the global asteroseismic parameters \dnu and \numax for a cohort of 173 stars observed across K2 campaigns 6-19, of which 159 are new detections. The sample mainly consists of MS dwarfs and subgiants but includes also a smaller set of low-luminosity RGs, and several known exoplanet hosts. We obtain a typical success rate in terms of seismic detections of ${\sim}50\%$ across campaigns. If we disregard the several proposed exoplanet hosts and cluster members with low expected detectability and the ones affected by an error in the calculation of K2 magnitudes which caused the downloaded pixel stamp to be too small to preserve the flux, the success rate is closer to ${\sim}63\%$ across the sample. Keeping in mind the prominent systematic noise source affecting K2 observations, and in turn the photometric quality, we consider this success rate to indicate that our selection strategy is robust and reliable.

We provide asteroseismic parameters from three independent pipelines and find a good consensus amongst these in terms of weighted RMS deviations at the level of ${\sim}2.5\%$ in \numax and ${\sim}0.35\%$ in \dnu, and with no indications of systematics. For the individual pipelines, we obtain typical fractional uncertainties of ${\sim}1.6\%$ in \numax and ${\sim}0.7\%$ in \dnu. The benefit of using several pipelines is evident from the different portions of the total sample identified as seismic sources by the different pipelines. Overall there are large overlaps with the majority of the sample identified by at least two independent pipelines.  

For the majority of the sample (163 out of 173) we obtain stellar atmospheric parameters homogeneously from spectroscopy with the SPC pipeline \citep{2012Natur.486..375B,Bieryla2024} on spectra from TRES. The spectroscopy is processed in an iterative manner in which the \logg was fixed to the asteroseismic one. This procedure is found to have a significant impact on the final results with systematic shifts in \teff by up to $\pm200$ K, in \logg by up to $\pm0.6$ dex, and in \feh by up to $\pm0.15$ dex. We find an excellent overall agreement between our spectroscopic results and those provided by several large spectroscopic surveys, including the GCS, APOGEE, LAMOST, and \textit{Gaia} for radial velocities. 

In addition to the spectroscopic parameters, we obtained \teff and angular diameters ($\theta$) from the IRFM \citep{Casagrande2021} for the majority of the sample. In the processing of these results, we test two different maps for the interstellar reddening and find that the \textit{Stilism} map \citep{Lallement2019}, as opposed to the \texttt{Bayestar19} map \citep{Green2019}, provides values that do not lead to a correlation between the reddening and the SPC-IRFM \teff difference and provide self-consistent reddening values for the stars of the M67 open cluster. Following the iteration of the spectroscopic analysis against the seismic \logg we find an excellent overall agreement between the two \teff-scales, in particular for the dwarfs and SGs, with only a minor systematic bias that leads to mean differences of the order ${\sim}20$ K at the limits of our \teff interval.

Our analysis shows the clear benefit of including several pipelines, both in terms of improving the yield of seismic detections and better assessing the systematic uncertainty of the seismic parameters. Similarly, the addition of different sources of information in the analysis of stellar atmospheric parameters has allowed us to reach a great consensus between the spectroscopic and IRFM \teff scales and again enables an assessment of the systematic uncertainty of the parameters.

We note that while we have focused on the global seismic parameters \dnu and \numax, a large portion of this new sample of seismic dwarfs and subgiants is amenable to a detailed analysis of individual modes of oscillation \citep{DaviesKAGES,legacy}, as evident from the example of \object{EPIC 212708252} shown in \fref{fig:data_ex}. Importantly, the stars of the KEYSTONE sample are typically significantly brighter than corresponding stars from the nominal \kp mission \citep{Mathur2021}, hence these will be more suitable for follow-up observations and characterisation from ground-based observations.
In a subsequent work, the sample of stars will undergo stellar modelling using the seismic and atmospheric parameters presented in this analysis.

\begin{acknowledgements}

The authors acknowledge the dedicated teams behind the \kp and K$2$ missions, without whom this work would not have been possible. Short-cadence data were obtained through the Cycle $1$-$6$ K$2$ Guest observer program (GO Program IDs: 1038, 2023, 3023, 4074, 5074, 6039, 7039, 8002, 10002, 11012, 12012, 13012, 14010, 15010, 16010, 17036, 18036, 19036), and associated NASA grants NNS16AE65G, NNX17AL49G, 80NSSC18K0363, and 80NSSC19K0102 to SB.
Funding for the Stellar Astrophysics Centre is provided by The Danish National Research Foundation (Grant agreement no.: DNRF106).
MNL acknowledges the support of the ESA PRODEX program.
DH acknowledges support from the Alfred P. Sloan Foundation and the Australian Research Council (FT200100871).
SH acknowledges support from the European Research Council via the ERC consolidator grant `DipolarSound' (grant agreement \#101000296).
TLC is supported by Funda\c c\~ao para a Ci\^encia e a Tecnologia (FCT) in the form of a work contract (CEECIND/00476/2018).
AMS acknowledges grants Spanish program Unidad de Excelencia Mar \'{i}a de Maeztu CEX2020-001058-M, 2021-SGR-1526 (Generalitat de Catalunya), and support from ChETEC-INFRA (EU project no. 101008324).
AS acknowledges support from the European Research Council Consolidator Grant funding scheme (project ASTEROCHRONOMETRY, G.A. n. 772293, \url{http://www.asterochronometry.eu}).
DS is supported by the Australian Research Council (DP190100666).
This work has made use of data from the European Space Agency (ESA) mission {\it Gaia} (\url{https://www.cosmos.esa.int/gaia}), processed by the {\it Gaia} Data Processing and Analysis Consortium (DPAC,
\url{https://www.cosmos.esa.int/web/gaia/dpac/consortium}). Funding for the DPAC has been provided by national institutions, in particular, the institutions participating in the {\it Gaia} Multilateral Agreement.\\

We acknowledge the use of the following Python-based software modules: \texttt{Astropy} \citep{Astropy}, \texttt{PyAstronomy} \citep{pya},  \texttt{Lightkurve} \citep{lightkurve}, \texttt{Emcee} \citep{emcee}, \texttt{PyMC3} \citep{pymc3},  \texttt{KDEpy} \citep{KDEpy}, \texttt{NumPyro} \citep{numpyro1,numpyro2}.

\end{acknowledgements}

\bibliographystyle{aa}
\bibliography{biblio}

\begin{appendix}

\section{SPC seismic \logg iteration}\label{sec:spclogg}

\begin{figure*}
    \centering
    \includegraphics[width=\textwidth]{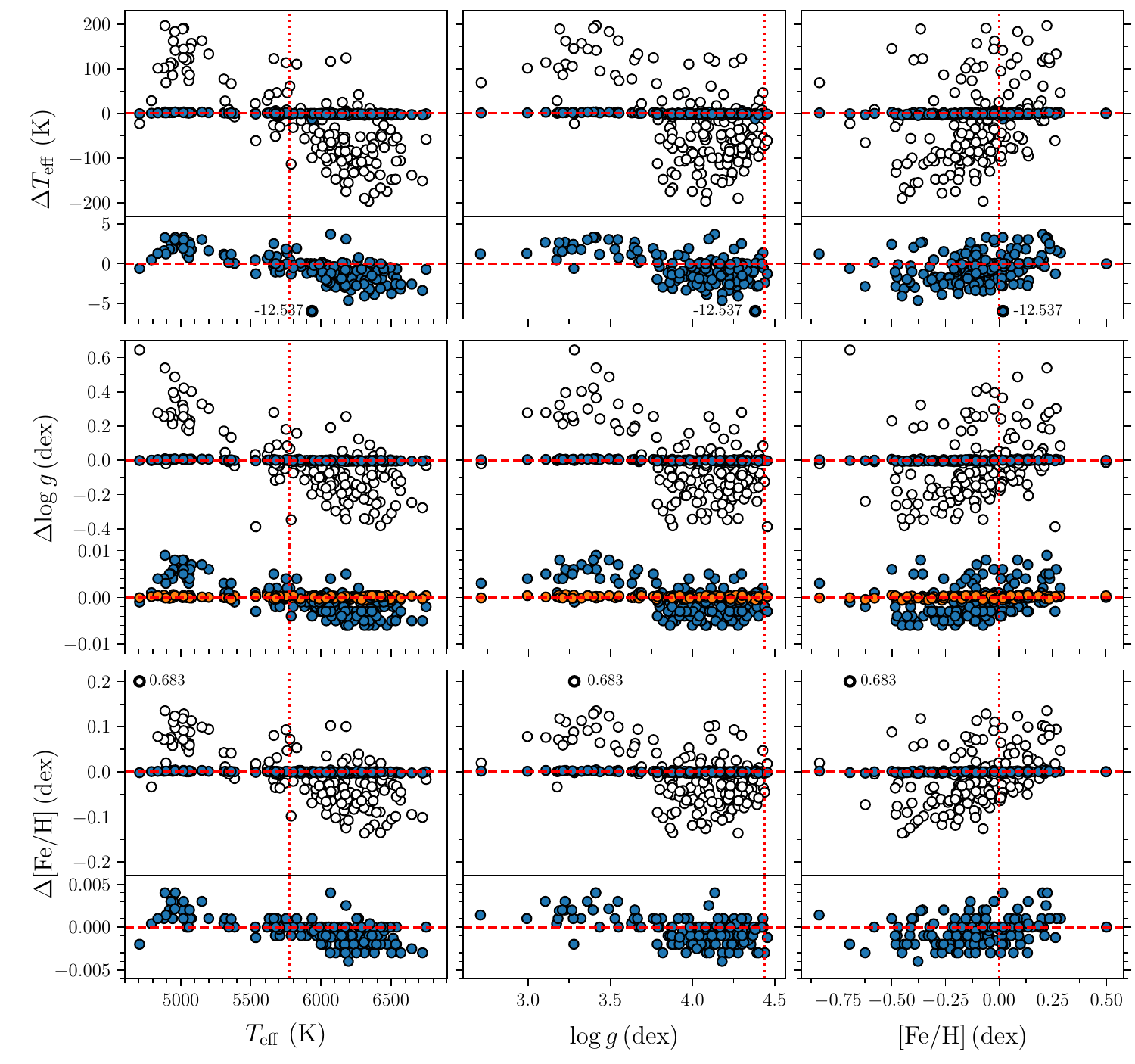}
    \caption{Change in \teff, \logg, and \feh from iterating the spectroscopic reduction with a \logg fixed to the seismic values obtained using \numax and \teff. Changes for a given parameter $X$ are given as $\Delta X = X_i - X_{i+1}$, with $i$ giving the step in the iteration. In all cases, changes are plotted against the adopted values after the second iteration. White markers indicate values from the first iteration, while blue markers indicate the second iteration. For all nine tiles, combining the changes in the three parameters with their corresponding values, the lower panel provides a zoomed version of the changed from the second iteration. For \logg the changes from a potential third iteration have been indicated by orange markers. The vertical red dotted lines show, respectively, the solar \teff, \logg, and \feh values for reference. In all panels, we have added a dashed red horizontal zero-change line. For $\Delta\teff$ and $\Delta\feh$ we have moved the ordinate position of two points for a better visualisation -- these have been indicated with a bold marker thickness and we have provided the actual value of the point.}
    \label{fig:spc_itt2}
\end{figure*}

We quantify in \fref{fig:spc_itt2} the dependence of the changes in the spectroscopic parameters caused by the iteration in the SPC analysis with the seismic \logg. The changes are shown as a function of \teff, \logg, and \feh, where the change for a given parameter $X$ is given as $\Delta X = X_i - X_{i+1}$, with $i$ giving the step in the iteration. A clear proportional dependence is seen, with a negative correlation of the parameter changes with \teff and \logg, and positive with \feh, and with the \teff and \feh dependence pivoting points around the solar values, while around a value of ${\sim}3.75$ dex for \logg. 
In terms of the parameter changes we find as expected strong correlations in the sense that changes in both \teff and \feh correlate positively with a change in \logg, with Pearson correlation coefficients in $\Delta\teff$ vs. $\Delta\logg$ and $\Delta\feh$ vs. $\Delta\logg$ of $\rho\sim0.93$ and $\rho\sim0.87$. In addition, we find indications of dependence on the correlation with \teff, in the sense that the correlation is stronger for stars with a \teff above the solar value. These dependencies, and changes from constraining \logg, are in good agreement with the findings of the dwarf sample of \citet{Torres2012} and exoplanet hosts sample of \citet{Huber2013}. We find no clear correlation between the change in the projected rotation velocity \vsini with \logg, but note a correlation with \teff (which is not surprising given the relationship between \teff and \vsini) with generally positive $\Delta(\vsini)$ values for $\teff \gtrsim 5750$ K (with an average ${\sim}0.1\,\rm km/s$) and larger negative values below this temperature (in the range ${-0.5}\,\, \rm to \,\, {-1.5}\,\rm km/s$).  

As seen from the bottom smaller panels in each of the tiles the change from the 1st to the 2nd iteration is small, and for the \logg values we further show the effect on this parameter from a potential 3rd iteration, which would result in insignificant changes, leading us to conclude the process after two iterations.

\section{Radial velocities and Doppler shifts}\label{sec:spcrv}

As a consistency check of the results from SPC we compare the measured RVs to those provided by \textit{Gaia} DR2 \citep{Soubiran2018} (these are also the ones adopted in \textit{Gaia} EDR3).
For SPC we use a correction for the Solar gravitational redshift of $\rm 0.61\, km/s$.  We note that twelve targets are missing RV values from \textit{Gaia} DR2.
\begin{figure*}
\centering
\includegraphics[width=.49\textwidth]{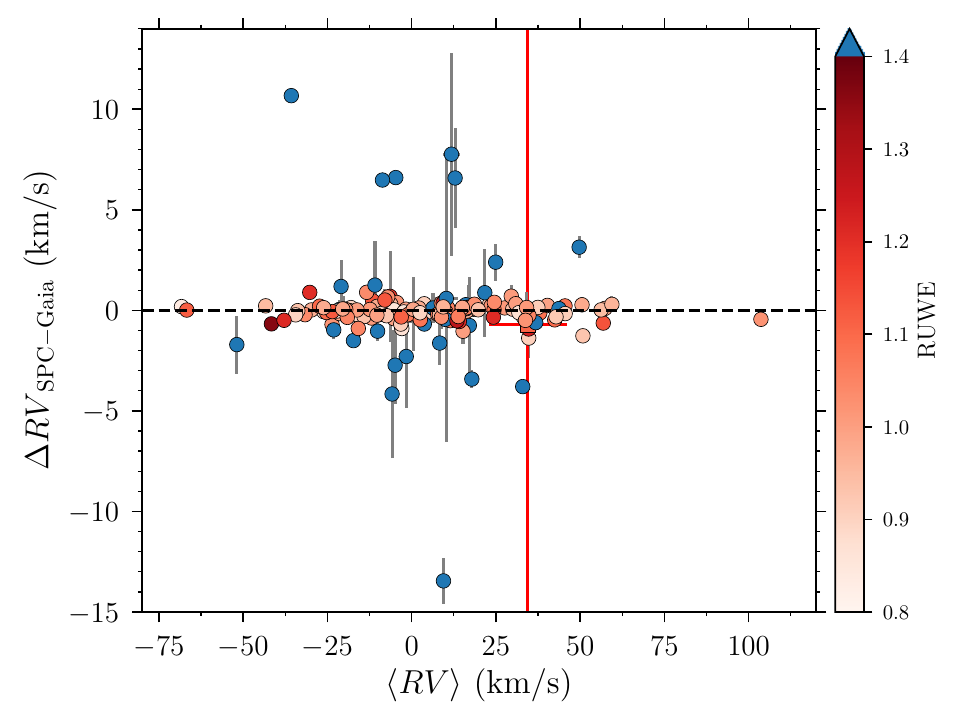}\quad\includegraphics[width=.49\textwidth]{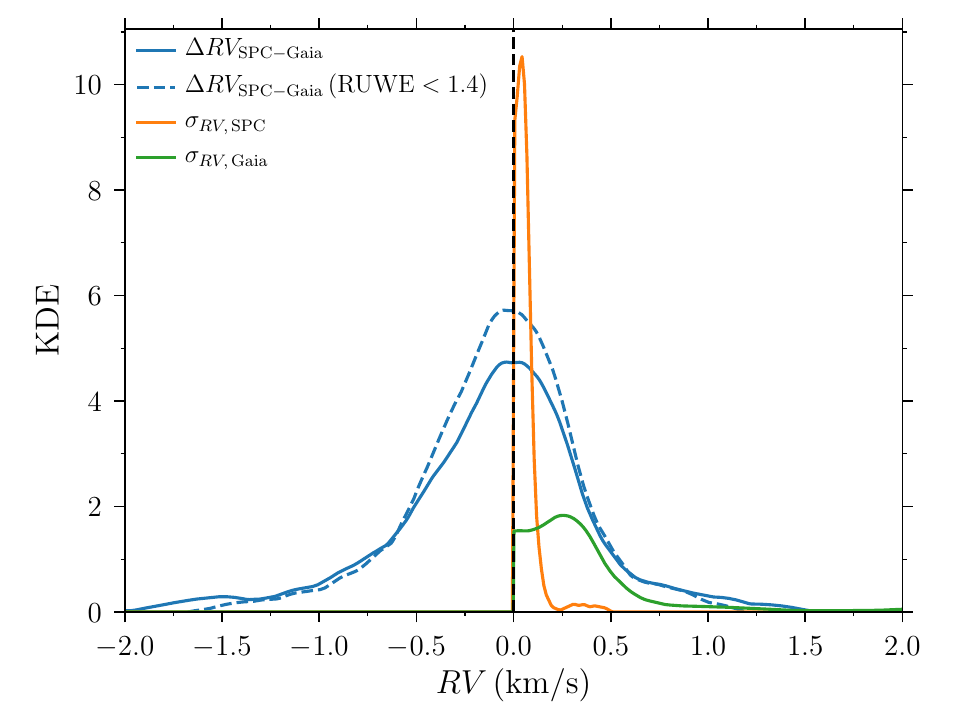}
\caption{Comparison between radial velocities from \textit{Gaia} DR2 and those from our SPC analysis. Left: difference in RV against the average, with the colours indicating the RUWE value from \textit{Gaia}. The target with red errorbars is the one $QF=4$ target with a \textit{Gaia} RV. Right: distributions of the RV differences (both including and excluding targets with high RUWE ($>1.4$) values), and RV uncertainties (see legend). For better visualisation we have increased the KDE values on the differences by a factor of 5.}\label{fig:rv}
\end{figure*}
The comparison of RVs is shown in \fref{fig:rv}. As seen the agreement is excellent, with a median and standardized MAD on the difference of only $\rm -0.036\, km/s$ and $\rm 0.4\, km/s$ and with no indication of proportional biases. As seen the largest differences are found for stars with a high RUWE ($>1.4$) value, indicating that the target is possibly non-single or otherwise problematic for the astrometric solution \citep{ruwe2018}. If we consider only low-RUWE targets the standardised MAD drops to $\rm 0.3\, km/s$, and with no differences beyond $\pm \rm 1.4\, km/s$. The median uncertainty on the \textit{Gaia} RVs of $\rm 0.27\, km/s$ nicely matches the scatter in the differences, where, by comparison, the median uncertainty on the SPC RVs is only at a level of $\rm 0.05\, km/s$.  

\begin{figure}
    \centering
    \includegraphics[width=0.5\textwidth]{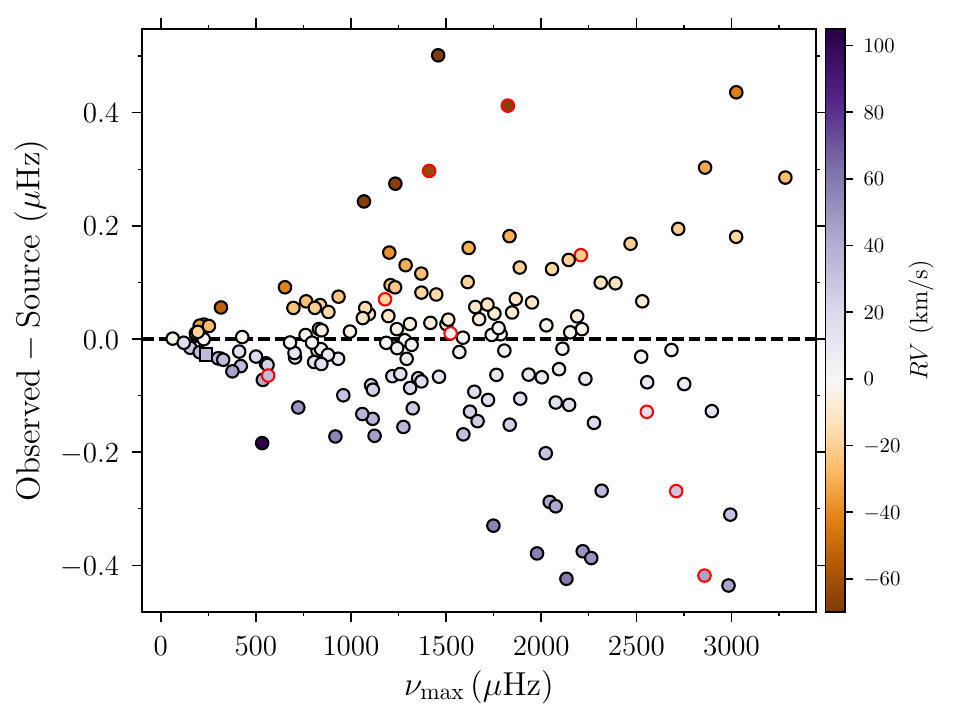}
    \caption{Effect from the Doppler shifts of observed oscillation modes from the stellar RV. The markers are colored according to the RV, and with black (red) outlines indicating RVs from SPC (\textit{Gaia} DR2).}
    \label{fig:dop}
\end{figure}

Following the prescription by \citet{Davies2014} we calculate the Doppler shift of the observed mode frequencies, hence \numax, from the stellar radial velocities. For nine of the ten stars without SPC results, we use RVs from \textit{Gaia} DR2 -- only for \object{EPIC 226083290} we lack a value for RV.
\fref{fig:dop} shows the resulting Doppler shifts from the RVs -- as seen the shifts are at maximum $\rm \pm 0.5\, \mu Hz$. Given the size of these shifts compared to the typical uncertainty on \numax of ${\sim}2\%$ (corresponding to $\rm {\sim}5.7\, \mu Hz$ for a $\rm \numax=283\, \mu Hz$) we choose to ignore this uncertainty contribution. However, for many of the stars in our sample peakbagging of individual oscillation modes is possible (\fref{fig:data_ex}), and here the Doppler shifts could in many cases be significant compared to the uncertainties on individual mode frequencies. 

\section{Reddening}\label{sec:app_red}

Initially, we adopted reddening values from the \citet{Green2019} extinction map \texttt{bayestar19}, using \textit{Gaia} EDR3 distances from \citet{BJ2021}. In 24 cases we obtained a non-zero $E(B-V)$, but still, many of these cases were tagged as unreliable by the map given the distance of the target. From these reddening values we noticed some significant outliers when comparing the \teff from the IRFM to those from spectroscopy, and a significant (negative) correlation in the \teff differences (SPC-IRFM) against $E(B-V)$, suggesting that the $E(B-V)$ values were overestimated. In addition, we found a large range in the extinction values ($E(B-V)$ from $0.02-0.08$ mag) for stars belonging to the M67 open cluster. This led us to consider the extinction map of the \textit{Stilism} project \citep{Lallement2014, Capitanio2017}. 
\begin{figure*}
    \centering
    \includegraphics[width=\textwidth]{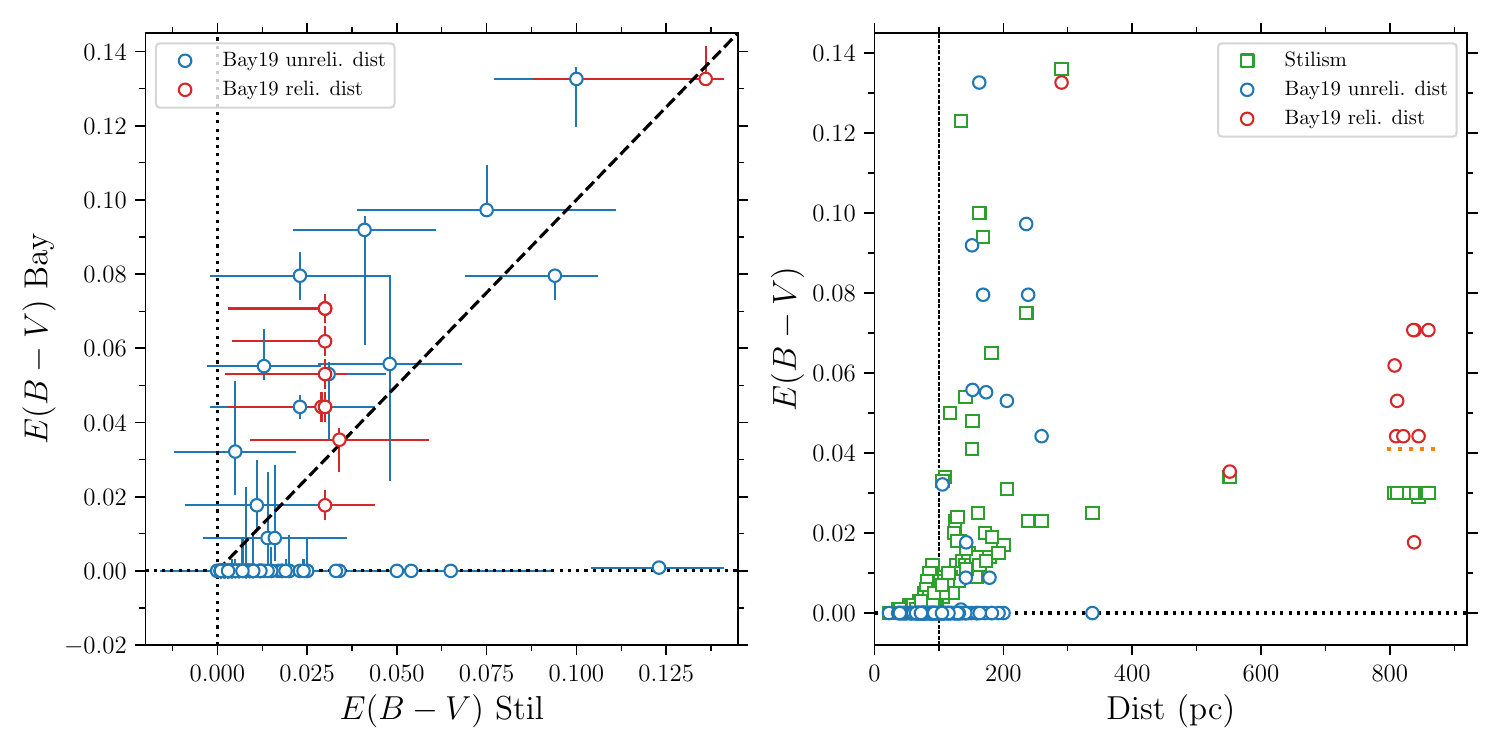}
    \caption{Comparison between reddening values from the  \texttt{bayestar19} \citep{Green2019} and the \texttt{stilism} maps \citep{Capitanio2017}. Left: direct comparison between $E(B-V)$ values from the two maps, including the $1:1$ correspondence by the dashed line. The marker colour indicates if the distance of the star was deemed within the reliable range in \texttt{bayestar19}. Right: $E(B-V)$ values as a function of distance. The dotted line at $100$ pc indicates the distance below which we adopt a zero reddening. The marker indicates the reddening source, and for the \texttt{bayestar19} values if $E(B-V)$ was deemed reliable given the distance. The orange horizontal dotted line at $E(B-V)=0.041$ mag and a distance $>800$ pc show the adopted reddening from \citet{Taylor2007} for M67 targets. }
    \label{fig:red_comp1}
\end{figure*}
\fref{fig:red_comp1} provides a comparison of the reddening values from the two sources, from which we see that (1) for the cases where both maps agree on a non-zero reddening the \texttt{bayestar19} values are generally larger than the \textit{Stilism} ones; (2) the \textit{Stilism} map nearly always return non-zero values, even in the near solar proximity -- we have chosen the approach of adopting a zero-reddening for stars closer than 100 pc; (3) the \textit{Stilism} map provides consistent $E(B-V)$ values from M67 stars (distance at $>800$ pc), though slightly lower than the adopted ones from \citet{Taylor2007}; (4) the reported uncertainties on the \textit{Stilism} values are a factor $4-5$ larger than the ones from \texttt{bayestar19} and likely overestimated -- we have adopted a $20\%$ uncertainty in estimating the impact on the derived IRFM \teff values.

From comparing the \teff differences between the IRFM and SPC from adopting the different reddening maps, we find that the \textit{Stilism} values reduce these and the correlation with the difference in $E(B-V)$, which for the \texttt{bayestar19} values could indicate a proportional bias.

\section{Comparison to other surveys}\label{sec:app_survey}

As a second consistency check of our spectroscopic SPC results, and to obtain metallicities for the IRFM derivation of \teff for the stars without SPC results, we make a comparison to some of the large spectroscopic surveys that overlap with our targets. Our comparison is made on stars in common with the Apache Point Observatory Galactic Evolution Experiment \citep[APOGEE;][]{apogee2020} DR16, The Radial Velocity Experiment \citep[RAVE;][]{RAVE2017} DR5, The Large Sky Area Multi-Object Fibre Spectroscopic Telescope \citep[LAMOST;][]{lamostK22020}, the GCS \citep[GCS;][]{2011A&A...530A.138C}, and GALactic Archaeology with HERMES \citep[GALAH][]{galah2021}.

For APOGEE and RAVE we make $S/N$-weighted average values when multiple spectra are available. For GCS no uncertainties are provided for \logg and \feh, so here we adopt uncertainties of $0.1$ dex, and a \teff uncertainty of $100$ K if no value is available.
\begin{figure*}
    \centering
    \includegraphics[width=\textwidth]{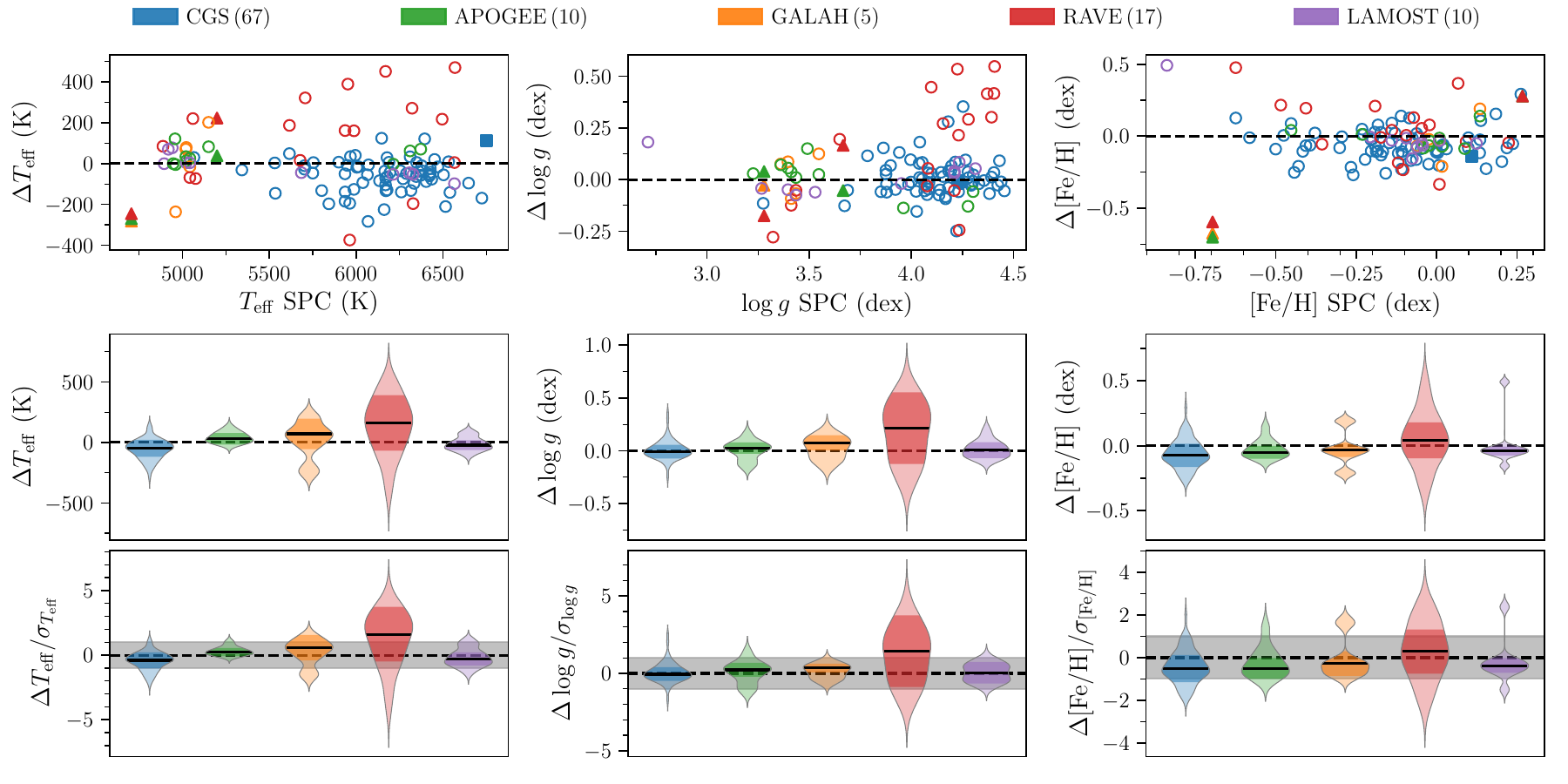}
    \caption{Comparison between our SPC results to those of spectroscopic surveys for \teff (left panels), \logg (middle panels), and \feh (right panels). The differences are given as $\rm \Delta X = X_{\rm SPC} - X_{\rm survey}$. The marker color indicates the comparison survey, see the top legend (the numbers in parenthesis indicate to number of stars in common with our SPC sample). Filled triangular (square) markers indicate the SPC $QF=4$ ($QF=3$) targets. Top row: value difference against SPC value. Middle row: violin plots showing the distributions of the differences, with the medians given by the solid black lines and the darker shaded regions giving the standardized MAD of the differences. Bottom row: Difference distributions, as in the middle row, but normalised by the uncertainty on the differences. The horizontal shaded region provides the $\pm 1 \sigma$ region. }
    \label{fig:surv}
\end{figure*}
\fref{fig:surv} shows the comparisons for \teff, \logg, and \feh. As seen, our SPC values in general agree well with the comparison surveys, with median differences (and scatter) within the uncertainty on the differences. The most significant disagreement is seen in the comparison with RAVE. Disregarding the RAVE values we see that for the $QF=4$ targets, the different surveys agree well with each other, but disagree with the SPC values. 
For the targets with no SPC results we opt for using values from APOGEE, as this survey has the largest overlap with this set of targets and as seen above generally agrees well with SPC.

\begin{figure}
    \centering
    \includegraphics[width=\columnwidth]{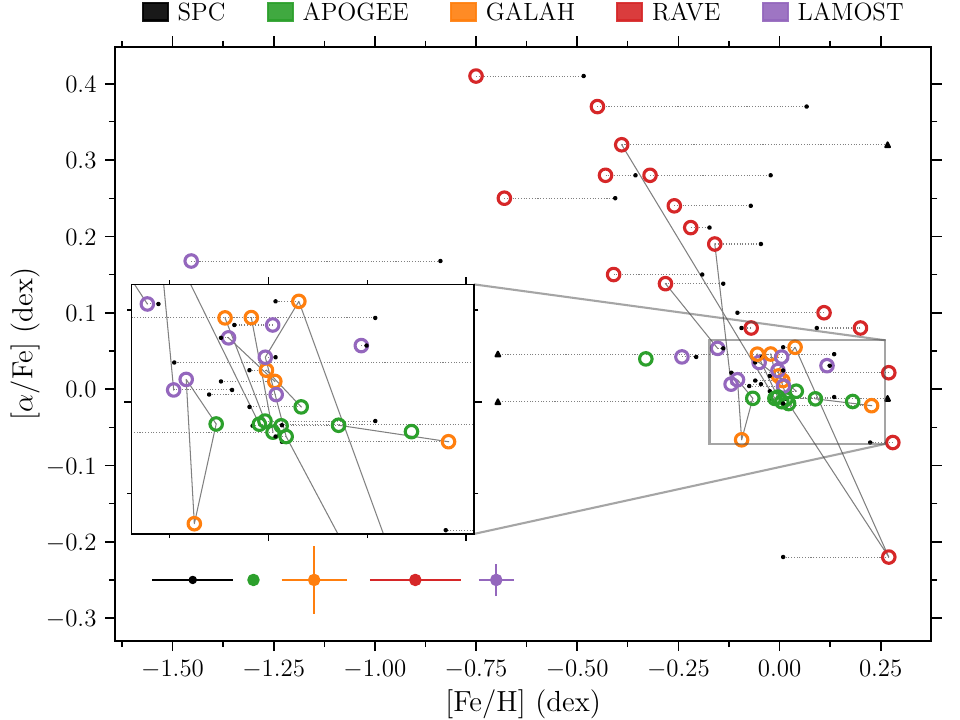}
    \caption{Metallicity (\feh) and $\alpha$-enhancement for stars included in the large spectroscopic surveys APOGEE, GALAH, RAVE, and LAMOST (see legend). \feh values from our SPC analysis are included if available, and connected to the corresponding survey value by a horizontal dotted line. Stars with values from several surveys are connected. Triangular SPC markers indicate the stars for which a poor spectroscopic analysis was obtained. The filled markers in the bottom left indicate the average uncertainties from the different sources. The insert shows a zoomed version of the most crowded region. }
    \label{fig:surv_alpha}
\end{figure}

We have also checked for the availability of $\alpha$-enhancements for our stars from the spectroscopic surveys. \fref{fig:surv_alpha} shows the available $\rm [\alpha/Fe]$ values against the corresponding \feh, and the value for \feh from our SPC analysis. As seen the $\rm [\alpha/Fe]$ values from the APOGEE, GALAH, and LAMOST surveys are, with a few exceptions, restricted to the interval $-0.025$ to $0.05$ dex, and with a good agreement (within errors) between there and our \feh values. While the surveys are all restricted to the above narrow interval, we note that in cases where several surveys provide values for the same star APOGEE typically provides a reduced $\rm [\alpha/Fe]$ close to a mean value of $\rm [\alpha/Fe]\approx -0.01$ dex. The values from RAVE are generally off from our \feh values (see also \fref{fig:surv}) and covering an extended region in $\rm [\alpha/Fe]$ with values in disagreement with the other surveys -- we therefore disregard values from RAVE in our analysis.


\section{Luminosity calculation}\label{app:lum}
For the luminosity calculations, we adopt values of $V_{\odot}$ mag and $BC_{V, \odot}$ mag from the analysis of empirical solar spectra, following the method of \citet{Casagrande2014}.

\renewcommand{\arraystretch}{1.1} 
\setlength\tabcolsep{0pt}
\begin{table}
\caption{Bolometric correction values.}
\label{tab:lum_bc}
\centering 
\begin{tabular*}{\linewidth}{@{\extracolsep{\fill}} lccc}
\toprule\addlinespace[3pt]
Ref. & $V_{\odot}$ & $M_{V,\odot}$ & $BC_{V,\odot}$ \\
\midrule\addlinespace[3pt]
HST/CALSPEC\tablefootmark{a} & -26.749 & 4.824 & -0.074 \\
\citet{Thuillier2004} & -26.749 & 4.823 & -0.073 \\
\citet{Rieke2008} & -26.736 & 4.836 & -0.086 \\
\citet{Meftah2021} & -26.743 & 4.829 & -0.079 \\
\bottomrule
\end{tabular*}
\tablefoot{$BC_{V,\odot}$ values are computed as $M_{\mathrm{bol},\odot} - M_{V,\odot}$ assuming $M_{\mathrm{bol},\odot}=4.75$.\\
\tablefoottext{a}{See \citet{Bohlin2014}}
}
\end{table}
\renewcommand{\arraystretch}{1} 

\tref{tab:lum_bc} lists the individual values from the four sources of empirical data considered, and we use the average values in our analysis. Our value for $V_{\odot}$ is in excellent agreement with the \citet{Torres2010} who lists $V_{\odot}=-26.76\pm 0.03$ mag, while $BC_{V, \odot}$ is slightly higher than the corresponding value from \citet{Casagrande2014} based on MARCS synthetic fluxes ($BC_{V, \odot}=-0.068\pm0.005$ mag, their table 2 using a microturbulent velocity of $\nu_{\rm micro}=2\, \mathrm{km/s}$ in the \texttt{VEGA} system). See also \citet{Torres2010} for an overview of previous empirical determinations.

For the extinction, computed as $A_{\xi} = R_{\xi} E(B-V)$, we use $R_{\xi}$ values from a \teff- and \feh-dependent relation similar to \citet{Casagrande2018}:
\renewcommand{\arraystretch}{1.1}
\setlength\tabcolsep{0pt}
\begin{table}
\caption{Extinction coefficients for \textit{Gaia} EDR3 filters.}
\label{tab:lum_r}
\centering
\begin{tabular*}{\linewidth}{@{\extracolsep{\fill}} ccccc}
\toprule\addlinespace[3pt]
Band & $a_0$ & $a_1$ & $a_2$ & $a_3$ \\
\midrule\addlinespace[3pt]
\multicolumn{5}{c}{\citet{Cardelli1989}} \\
\midrule\addlinespace[3pt]
$G$ & 1.472 & 2.931 & -1.393 & -0.011 \\
$BP$ & 2.045 & 3.392 & -2.067 & -0.022 \\
$RP$ & 1.818 & 0.456 & -0.206 & 0.003 \\
\midrule\addlinespace[3pt]
\multicolumn{5}{c}{\citet{Fitzpatrick1999}} \\
\midrule\addlinespace[3pt]
$G$ & 1.132 &  2.700 & -1.271 & -0.010 \\
$BP$ & 1.684 &  3.098 & -1.879 & -0.020 \\
$RP$ & 1.471 &  0.369 & -0.167 &  0.002 \\
\bottomrule
\end{tabular*}
\tablefoot{Extinction coefficients to be used in Eq.~\ref{eq:rval}, with values for both the \citet{Cardelli1989} and \citet{Fitzpatrick1999} (renormalised as per \citet{Schlafly2016}) extinction laws.
}
\end{table}
\renewcommand{\arraystretch}{1}

\begin{equation}\label{eq:rval}
    R_{\xi} = a_{0,\xi} + \mathrm{T4}(a_{1,\xi} + a_{2,\xi}\mathrm{T4}) + a_{3,\xi}\feh\,\, ,
\end{equation}
where $\mathrm{T4} = \teff/1\mathrm{e}4$, and use revised coefficients suitable to \textit{Gaia} EDR3. The coefficients entering \eqref{eq:rval} are provided in \tref{tab:lum_r} for \textit{Gaia} EDR3 $G$, $BP$, and $RP$ bands, and for both the \citet{Cardelli1989} and \citet{Fitzpatrick1999} (renormalised as per \citet{Schlafly2016}) extinction laws. 

\section{Comparisons and checks of global seismic parameters}\label{sec:app1}

\subsection{Selection strategy evaluation}
\begin{figure*}
    \centering
    \includegraphics[width=\textwidth]{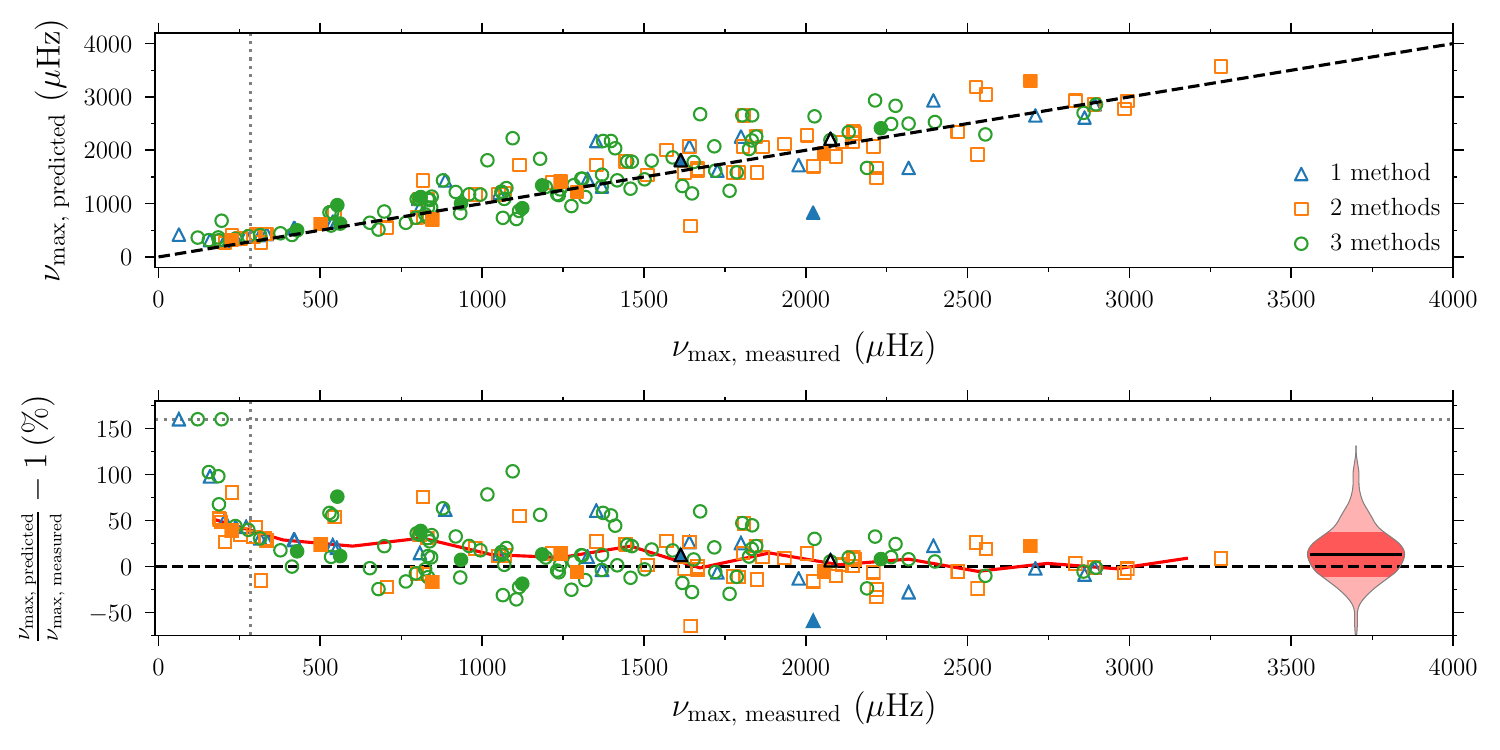}
    \caption{Comparison between measured and predicted values for \numax for stars in C11-19 (empty markers) and and C8+10 (filled markers), with measured values from the \texttt{CV} method. The two targets with black edges indicate the targets where only the \texttt{SYD} method returned detections. Top: a direct comparison, where the dashed line gives the $1:1$ relation. The dotted vertical line indicated the Nyquist frequency of $\rm {\sim}283\, \mu Hz$ for $30$-min long-cadence data (also shown in the bottom panel). Bottom: relative differences between measured and predicted values against the measured values. Targets with a fractional difference above $160\%$ have been moved to this value, as indicated by the dotted horizontal line. The full red line connects ten median-binned values across the \numax range. To the right in this panel, we show a violin plot of the distribution, with the median indicated by the full black line and the spread indicated by the darker shaded interval. The marker type/colour indicates the number of methods for extracting seismic parameters that agree with a positive detection (see legend).}
    \label{fig:preest}
\end{figure*}

As an evaluation of our target selection methodology, we compare the measured values of \numax to those predicted for the target selection. The comparison is shown in \fref{fig:preest} for the stars observed in C11-19, where the detectability calculation of \citet{keystone} was used, and C8+10 using the version of \citet{2011ApJ...732...54C} \citep[see also][]{k2chap}. The median offset is of the order ${\sim}10\%$, and with a spread of ${\sim}25\%$ which is to be expected given the many assumptions and parameters entering the detectability calculation, all of which have their own sources of uncertainty. There is a slight systematic trend in the differences with \numax being increasingly over-predicted towards lower measured \numax -- towards the Solar \numax (${\sim}3090\, \rm \mu Hz$) the offset and scatter decreases, which might be expected given the frequent referencing to the Sun in the various scaling relations entering the \numax prediction. 

Concerning the success rate in the number of detections as a function of magnitude, we find a fairly stable return of the order ${\sim}60\%$ up until $\Kp{\sim}9.5$. Beyond this magnitude, the success rate is lower, but there are also fewer proposed stars here, many of which are either suspected members of open clusters or exoplanet candidate host stars, hence proposed with a known lower predicted detectability.  

\subsection{Multi-campaign targets}
\begin{figure*}
    \centering
    \includegraphics[width=\textwidth]{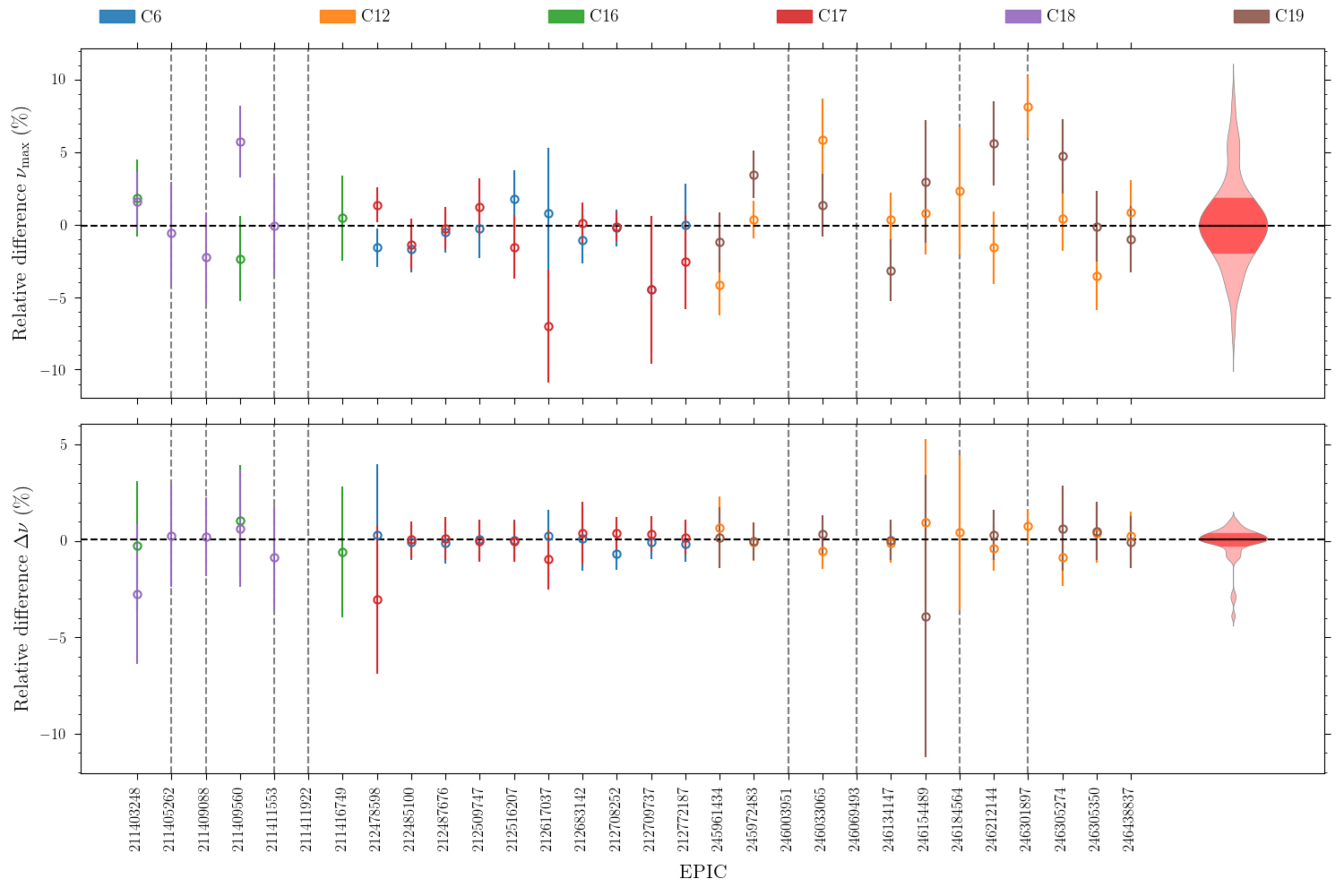}
    \caption{Comparison for multi-campaign targets of measured \numax (top) and \dnu (bottom) values from individual campaigns as compared to the value from the joint data. The values shown are based on the \texttt{CV} method. The colour indicates the campaign, and the horizontal dashed line indicates the median of the differences -- the violin plot to the right shows the distribution of differences and indicates the $1-\sigma$ spread (given by the standardised MAD). Vertical dashed lines indicate stars for which a detection was not obtained from a given or any of the single campaigns, but in all cases was obtained from the joint data. }
    \label{fig:av_multi_comp}
\end{figure*}

\fref{fig:av_multi_comp} shows the comparison of global seismic parameters obtained with the \texttt{CV} from individual campaigns to those obtained from combining the campaigns. We note that for M67 targets, the combined data also included data from C5, while no individual campaign estimates were obtained for C5 (for this reason, 211416749 only has a C16 value and a joint value, and no vertical dashed line). We find in general an excellent agreement from individual and combined data values. Similar levels of agreement are obtained from the \texttt{SYD} and \texttt{TACO/OCT} methods (not shown), though for fewer stars than the \texttt{CV} method.

\end{appendix}

\end{document}